\documentclass[10pt]{article}
\usepackage[accepted]{tmlr}

\usepackage{amsmath}
\usepackage{amssymb}
\usepackage{mathrsfs}
\usepackage{bm}

\usepackage{hyperref}
\usepackage[hypcap=false]{caption}

\usepackage{graphicx}
\usepackage{longtable}
\usepackage{booktabs}
\usepackage{array}
\usepackage{algorithm}
\usepackage{algpseudocode}
\usepackage{url}
\usepackage{booktabs}
\usepackage{adjustbox}
\usepackage{wrapfig} 

\DeclareMathOperator*{\argmax}{arg\,max}

\usepackage{mathtools}
\usepackage{amsthm}

\usepackage{tikz}
\usepackage{pgfplots}
\pgfplotsset{compat=1.18}

\usetikzlibrary{arrows.meta, decorations.markings}

\usepackage{multirow} 
\usepackage{siunitx}
\usepackage[T1]{fontenc}
\usepackage{lmodern} 
\usepackage{caption}
\usepackage{float}

\usepackage{placeins}

\title{NSFL: A Post-Training Neuro-Symbolic Fuzzy Logic (NSFL) $\Delta$ Framework for Boolean Operators in Neural Embeddings }

\author{%
  \name Vladi Vexler\thanks{Corresponding author. \texttt{vladi.vexler@huawei.com}},
        Ofer Idan, Gil Lederman, Dima Sivov \\
  \addr Huawei Tel-Aviv Research Center
}

\def\month{10}
\def\year{2026}
\def\openreview{\url{https://openreview.net/forum?id=WYpsWBabCu}}
\begin{document}

\maketitle
\begin{abstract}
Standard dense retrievers have no native mechanism for multi-atom logical
constraints. We introduce \textbf{Neuro-Symbolic Fuzzy Logic (NSFL)}, a
training-free framework that brings fuzzy-logical structure to neural
embedding spaces through a first-order finite-difference treatment of
similarity. NSFL anchors logical operations on zero-order atomic scores while
steering with \textbf{Neuro-Symbolic Deltas (NS-$\Delta$)} --- the marginal
semantic contributions induced by contextual fusion --- preserving atomic
meaning while capturing contextual dependence, and avoiding the representation
collapse and manifold escape endemic to geometric baselines. For
latency-constrained deployment, \textbf{Spherical Query Optimization (SQO)}
uses Riemannian optimization to project logical formulas into manifold-stable
query vectors. Across six encoder configurations and two modalities, NSFL
yields mAP@100 improvements up to \textbf{+81\%} on QUEST and \textbf{+70\%}
on a COCO-derived text-to-image benchmark, and provides an additive
\textbf{+19\% average} gain on an encoder already fine-tuned for logical
retrieval. On NevIR, where absolute accuracies are directly interpretable,
NSFL raises pairwise accuracy, NevIR’s both-correct criterion, from \textbf{0.256 to 0.508}. The
decomposition is exact and general; the operators built on it are engineering
realizations rather than a closed algebra, and we state that boundary precisely.
Code is available at \url{https://github.com/vladiaidev-lab/NSFL}.
\end{abstract}

\section{Introduction}
Neural embeddings transform raw data into high-dimensional dense vectors,
capturing semantic relationships for tasks like retrieval~\citep{reimers2019sentence,
radford2021learning}. However, these embeddings struggle with Boolean logic
queries (e.g., ``A AND NOT B''), as fusions often yield additive effects
rather than crisp set operations. While much work on neuro-symbolic
integration operates during training~\citep{badreddine2022logic}, the
Neuro-Symbolic Fuzzy Logic (NSFL) framework operates post-training. We
motivate NSFL through empirical analysis of how encoders behave under logical
fusion, and build operators on a first-order finite-difference treatment of
similarity.

\begin{figure}[ht]
    \centering
    \begin{tikzpicture}[scale=1.8, >=Stealth]
        
        \tikzset{
            manifold/.style={blue!10, opacity=0.6},
            sphere_bg/.style={gray!20, fill=gray!5},
            vector_main/.style={->, thick, gray!60},
            vector_result/.style={->, ultra thick},
            label_style/.style={scale=0.8, align=center}
        }

        \begin{scope}[shift={(0,0)}]
            \draw[sphere_bg] (0,0) circle (1);
            \draw[gray!40, dashed] (1,0) arc (0:180:1 and 0.3);
            
            \fill[manifold] (-0.15,0.95) to[bend left=15] (0.9,0.2) -- (0.7,0.0) to[bend right=15] (-0.2,0.75) -- cycle;

            \draw[vector_main] (0,0) -- (0.1, 0.95) node[above, black, scale=0.8] {$A$};
            \draw[vector_main] (0,0) -- (0.95, 0.1) node[right, black, scale=0.8] {$B$};
            
            \draw[vector_result, red] (0,0) -- (-0.5, 0.6) node[left, scale=0.8] {$A-B$};
            
            \node[red, label_style, anchor=south] at (-0.8, 0.8) {\textbf{Manifold Escape}};

            \node at (0,-1.3) {\small (a) Negation by Linear Subtraction};
        \end{scope}

       \begin{scope}[shift={(2.8,0)}]
            \draw[sphere_bg] (0,0) circle (1);
            \draw[gray!40, dashed] (1,0) arc (0:180:1 and 0.3);
            
            \fill[manifold] (-0.15,0.95) to[bend left=15] (0.9,0.2) -- (0.7,0.0) to[bend right=15] (-0.2,0.75) -- cycle;
            
            \draw[vector_main, gray!60] (0,0) -- (0.1, 0.95) node[above, gray!60, scale=0.8] {$A$};
            \draw[vector_main, gray!60] (0,0) -- (0.95, 0.1) node[right, gray!60, scale=0.8] {$B$};
            
            \draw[->, orange!90!red, thick] (0,0) -- (0.5, 0.5) node[above right, orange!90!red, scale=0.7]  {$A$ not $B$}; 
            
            \coordinate (v_not) at (-0.02, 0.95);
            \draw[vector_result, green!60!black, ultra thick] (0,0) -- (v_not);
            
            \node[left=4pt, green!60!black, scale=0.9, font=\bfseries] at (v_not) {$x^*_{A \land \neg B}$};
            
            \draw[->, dashed, gray, out=30, in=150] (0.5, 0.5) to (v_not);
            
            \node[green!60!black, label_style, anchor=south] at (0, 1.1) {\textbf{Inhibitory Steering}\\(Preserves $A$, Excludes $B$)};
        
            \node at (0,-1.3) {\small (b) NSFL: $A \text{ NOT } B$};
        \end{scope}

        \begin{scope}[shift={(5.6,0)}]
            \draw[sphere_bg] (0,0) circle (1);
            \draw[gray!40, dashed] (1,0) arc (0:180:1 and 0.3);
            
            \fill[manifold] (0.2,0.9) to[bend left=15] (0.9,0.2) -- (0.7,0.0) to[bend right=15] (0.0,0.7) -- cycle;
            
            \draw[vector_main] (0,0) -- (0.1, 0.95) node[above, black, scale=0.8] {$A$};
            \draw[vector_main] (0,0) -- (0.95, 0.1) node[right, black, scale=0.8] {$B$ (Weak)};
            
            \coordinate (mono) at (0.25, 0.8);
            \draw[->, thick, orange] (0,0) -- (mono) node[right, scale=0.7] {$AB$};
            
            \coordinate (v_and) at (0.5, 0.65);
            \draw[vector_result, blue!80!black] (0,0) -- (v_and) node[right, blue!80!black, scale=0.9] {$x^*_{A \land B}$};
            \draw[->, dashed, gray] (mono) to (v_and);
            
            \node[blue!80!black, label_style, anchor=south] at (0.6, 1.1) {\textbf{Balanced Conjunction}\\(Pull to Weak Signal)};

            \node at (0,-1.3) {\small (c) NSFL: $A \text{ AND } B$};
        \end{scope}
        
    \end{tikzpicture}
    \caption{Comparison of embedding manifold operations: (a) Euclidean operations cause \textbf{Manifold Escape} and signal loss; (b) NSFL Inhibitory steering preserves context $A$ while excluding $B$; (c) NSFL Conjunctive steering resolves the \textbf{Semantic Bottleneck} by pulling the query toward the weaker atom $B$.}
    \label{fig:manifold_steering_comparison}
\end{figure}
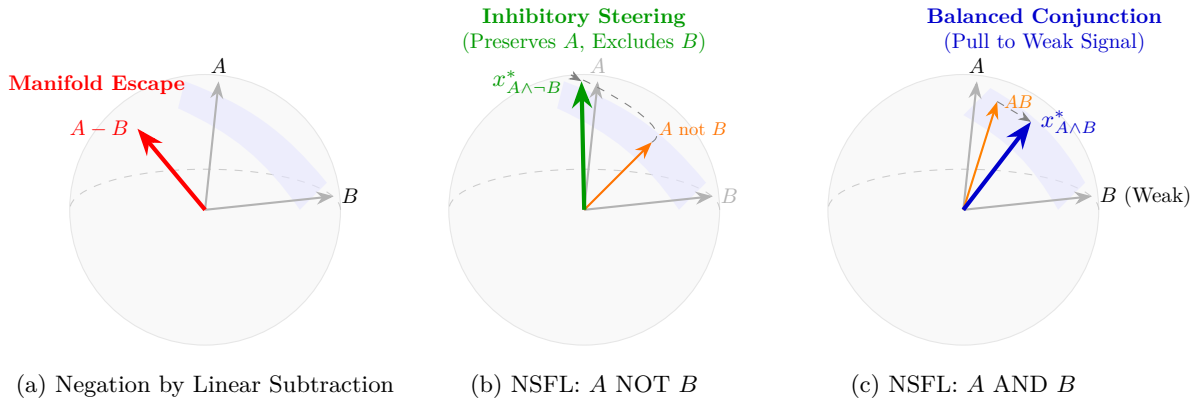

\paragraph{Why Boolean failure matters in practice.}
Logical constraints are how users express exclusion and refinement, not an
edge case: e-commerce search receives ``running shoes but not Nike'', patent
and legal discovery turns on exclusionary claims, clinical literature search
narrows by contraindication, and retrieval-augmented generation issues
filtered subqueries whose intent must survive the embedding step. The failure
mode is costly because it is silent --- a retriever that treats ``$A$ but not
 $B$'' as ``$A$ and $B$'' returns results that are confidently wrong, with
nothing downstream to signal the error. Since such systems are typically built
on frozen encoders and pre-built indices, the practical question is not
whether a differently trained retriever could handle logic, but whether
logical intent can be recovered \emph{post hoc} from an existing one.

To this end, we summarize the primary contributions of our work as follows:
\begin{itemize}
\item \textbf{A First-Order Semantic Decomposition:} We introduce the
\textbf{Neuro-Symbolic Delta ($\Delta$)}, an exact first-order
finite-difference decomposition of a fused similarity score into its atoms'
semantic contributions. This shifts dense retrieval from zero-order similarity
points to first-order differences, isolating individual atomic contributions
within a fused representation and avoiding the compensation inherent to convex
aggregation.

\item \textbf{Manifold-Aware Logical Operators:} We define asymmetric
Conjunctive (AND), Inhibitory (NOT), and Disjunctive (OR) operators that steer
queries without signal collapse. Conjunction inherits the non-compensatory
principle distinguishing t-norms from averaging operators; disjunction extends
the Zadeh t-conorm to incorporate fused and monolithic scores, accommodating
the superadditivity common in neural retrievers; and inhibition departs from
fuzzy negation deliberately, since encoders preserve rather than invert
semantic direction. Each carries a \textbf{boundary stability} condition that
defers to the encoder in regimes of extreme reinforcement or rejection.

\item \textbf{Deployment via Spherical Query Optimization (SQO):} We introduce
SQO, which projects multi-atom logical constraints into a single
manifold-stable query vector by Riemannian optimization. SQO improves over
monolithic retrieval at standard ANNS latency without post-retrieval
computation, though reranking subsumes its gains where it is affordable. We
characterize which operator components a pre-retrieval objective can express
and which it structurally cannot.

\item \textbf{Robust Empirical Generalization:} We validate our framework
across six encoder configurations spanning vision and text modalities.
NSFL consistently outperforms monolithic querying and geometric baselines,
with mAP improvements up to \textbf{+81\%}, and outperforms classical
score-combination baselines including a calibrated variant. Applied to
LogiCoL~\citep{shen2025logicol}, it yields a further \textbf{+19\% average
improvement}, demonstrating additive gains on top of training-time methods.
\end{itemize}

\section{Related Work}
\label{sec:related_work}

The integration of symbolic logic with neural representations has evolved from early fuzzy systems to modern high-dimensional dense retrievers. Our work sits at the intersection of fuzzy set theory, vector-symbolic architectures, and modern representation learning.

\subsection{Classical Fuzzy Logic and Multi-Valued Logic}
Modern fuzzy logic traces its lineage to \citet{zadeh1965fuzzy, zadeh1975fuzzy}, who introduced fuzzy sets and their subsequent logical formalizations. Classical operators rely on t-norms and t-conorms, such as those utilized in the Mamdani \citep{mamdani1975advances} and Sugeno \citep{sugeno1974theory} inference systems. These foundations were later expanded into continuous relaxations of Boolean operators to handle uncertainty in symbolic domains \citep{hajek2001metamathematics}.

\paragraph{Limitations:} A primary bottleneck when applying classical fuzzy logic to neural latent spaces is the assumption of calibrated probabilistic scores. As noted by \citet{guo2020deep}, neural ranking models are typically optimized for relative ranking via contrastive losses rather than absolute truth values. This leads to uncalibrated similarity distributions where standard t-norm axioms, such as monotonicity, are frequently violated.

\subsection{Vector Symbolic Architectures and Compositionality}
Structural compositionality was explored extensively via Vector Symbolic Architectures (VSA) before the deep learning era. Notable frameworks include Tensor Product Representations \citep{smolensky1990tensor} and Sparse Distributed Memory \citep{kanerva1988sparse}. Contributions such as Holographic Reduced Representations \citep{plate1995holographic, plate2003holographic} and Multiplicative Binding \citep{gayler1998multiplicative} provided algebraic pathways for logic-like operations in vector spaces.

\paragraph{Limitations:} VSAs require explicit binding operations (e.g., circular convolution) that are not natively supported by standard contrastive training objectives, such as CLIP \citep{radford2021learning} or SigLIP \citep{zhai2023sigmoid}, which do not inherently support the specific binding operations required by VSA.

\subsection{In-Training Neuro-Symbolic Models}
Significant progress has been made in differentiable logic and end-to-end neuro-symbolic integration. Logic Tensor Networks (LTN) \citep{badreddine2022logic, serafini2016logic} and DeepProbLog \citep{manhaeve2018deepproblog} represent the state-of-the-art in learnable logical constraints. Other frameworks, such as the Neuro-Symbolic Concept Learner \citep{mao2019neuro} and differentiable proving systems \citep{rocktaschel2017end}, focus on grounding symbolic reasoning within neural architectures.

\paragraph{Limitations:} These methods typically require full optimization or specialized training regimes. As discussed by \citet{garcez2022neurosymbolic}, training-based Neuro-Symbolic AI is powerful but computationally intensive. In contrast, post-training interventions are often more practical for large-scale, pre-trained retrieval systems.

\subsection{Logic in Modern Dense Retrievers}
Recent benchmarks such as QUEST \citep{malaviya2023quest}, NevIR \citep{weller2024nevir}
, and FSIR-BD \citep{idan2026fsir}  
have exposed the logical deficiencies of standard encoders like SBERT \citep{reimers2019sentence}. Research has responded with query decomposition \citep{geva2021did}, geometric region-based approaches such as Box \citep{hamilton2018embedding, chen2020probabilistic} and Beta Embeddings \citep{ren2020beta}, and logic-aware contrastive learning frameworks like LogiCoL \citep{shen2025logicol}. As empirically demonstrated by \citet{shen2025logicol}, dense retrievers frequently suffer from representation collapse, treating logically contradictory queries (e.g., ``A AND B'' versus ``A NOT B'') as highly correlated semantic equivalents ($r \approx 0.88$). However, correcting this via LogiCoL requires computationally expensive in-training t-norm regularizations on specifically curated logical mini-batches. Furthermore, even with advanced training regimes, modern retrievers face fundamental geometric hurdles: \citet{weller2025theoretical} established that embedding dimensionality imposes a strict theoretical ceiling on the number of document subsets a single query vector can retrieve, while most existing multi-atom fusion methods rely on convex aggregation, which \citet{grabisch1996equivalent} identifies as being strictly limited by the range of the constituent inputs.

\paragraph{Set-Based and Geometric Query Embeddings.}
A parallel line of research represents logical queries as geometric regions 
rather than point vectors, such as Query2Box~\citep{ren2020query2box} and 
ConE~\citep{zhang2021cone}. While effective for knowledge graph completion, 
these methods require specialized training objectives and custom index 
structures incompatible with standard ANNS pipelines. NSFL, by contrast, 
operates post-hoc on arbitrary pre-trained encoders using standard cosine 
similarity, requiring no architectural modification or retraining.

\paragraph{Comparison with NSFL} 
NSFL departs from existing logical retrieval frameworks by transitioning from zero-order similarity to a training-free, first-order finite-difference treatment. While prior methods focus on embedding-space reorganization, NSFL distinguishes itself through three core architectural advantages:

\begin{itemize}
    \item \textbf{Post-hoc Conflict Resolution:} NSFL resolves logical conflicts post-hoc, avoiding the global latent space reorganization required by in-training regularizations like LogiCoL, preventing degradation of the encoder's general retrieval performance.
    \item \textbf{Non-convex Steering via the NS-Delta ($\Delta$):} By extracting the Neuro-Symbolic Delta, NSFL enables non-convex steering that bypasses the capacity constraints identified by \citet{weller2025theoretical}, ensuring atomic logical contributions are neither diluted nor lost.
    \item \textbf{Orthogonality to SOTA:} NSFL is complementary to training-time optimizations. Applying NSFL as a reranking layer on top of \texttt{LogiCoL-E5} yields a further \textbf{19\% average mAP improvement}, demonstrating that algebraic reranking captures nuances that training-time signals alone may miss.
\end{itemize}

\section{Empirical Observations on Embedding Behaviors}
\label{sec:empirical_observations}

Our framework is grounded in observed geometric behaviors across six modern contrastive encoder configurations, including BLIP, SigLIP, BGE-Large v1.5, Nomic Embed Text v2, \texttt{e5-base-v2}, and the logic-fine-tuned \texttt{LogiCoL-E5v2}. Despite their heterogeneous architectures, these models exhibit consistent geometric patterns highlighting the divergence between dense vector similarity and formal Boolean logic.

\begin{itemize}
    \item \textbf{Non-probabilistic and Uncalibrated Scores:} Similarity scores in dense retrieval are typically uncalibrated and optimized for relative ranking rather than probability estimation \citep{guo2020deep}. Standard retrievers often collapse disparate logical operators into a single ``union-like'' representation, failing to distinguish between intersection and negation \citep{shen2025logicol}.
    
    \item \textbf{The State Space of Semantic Aggregation:} Unlike classical fuzzy logic which enforces strict monotonicity ($a \land b \leq \min(a,b)$), concept fusion in neural latent spaces is highly non-linear. When atomic concepts ($A$ and $B$) are fused into a joint query representation ($A \cup B$), the resulting similarity score ($S_{AB}$) occupies a state space characterized by three behaviors:
    \begin{enumerate}
        \item \textit{Superadditivity:} $S_{AB} > S_A + S_B$ (synergistic amplification or ``co-occurrence reward'' \citep{yuksekgonul2023when}).
        \item \textit{Subadditivity:} $S_{AB} < S_A + S_B$ while $S_{AB} \geq \min(S_A, S_B)$.
        \item \textit{Strong Decrease:} $S_{AB} < \min(S_A, S_B)$ (reductive interference \citep{weller2024nevir}).
    \end{enumerate}
    NSFL leverages this state space post-hoc, utilizing first-order delta analysis to map these fluctuations to precise logical operations.
    
    \item \textbf{Negation Stability (Negation as Noise):} Probing tasks reveal that negation tokens (e.g., ``not'') are often treated as semi-meaningful noise that fails to induce semantic reversal \citep{weller2024nevir, kassner2020negated}. Our measurements (Appendix~\ref{app:principles}) confirm this directional preservation in 69--78\% of candidates across encoders. This lack of inversion necessitates a ``positive-only'' operator approach to avoid instability in the negative regime of the manifold \citep{thrush2022winoground}.
\end{itemize}

\section{NSFL Empirico-Conjectural Framework}
\label{sec:nsfl-conjecture}

Following the principles of representation learning~\citep{bengio2013representation, lecun2015deep}, we propose an \textbf{Empirico-Conjectural framework} grounded in the geometric behaviors observed in Section~\ref{sec:empirical_observations}.

\subsection{Notation and Preliminaries}
\label{subsec:definitions}

\begin{table}[ht]
\centering
\caption{Summary of Notation and Definitions.}
\label{tab:notation}
\begin{small}
\begin{tabular}{@{}ll@{}}
\toprule
\textbf{Symbol} & \textbf{Description} \\ \midrule
 $\mathcal{D} = \{D_1, \dots, D_m\}$ & A corpus of data items embedded as vectors $\mathbf{v}_{D_i} \in \mathbb{R}^d$. \\
 $L_A, L_B, L_C$ & Semantic query atoms (e.g., text labels or raw concepts). \\
 $\cup$ & The Fusion Operator (FO): $L_{AB} = L_A \cup L_B$. Merges atoms into a joint label. \\
 $L_M$ & Monolithic Query: The full boolean expression as a single natural language string. \\
 $\mathbf{v}_X$ & Embedding vector of query component $X \in \{A, B, AB, M\}$. \\
 $S_X^i$ & Similarity score between data item $i$ and query $X$, normalized to $[0, 1]$. \\
 $\Delta_X$ & \textbf{Neuro-Symbolic Delta}: The marginal change $S_{AX}^i - S_A^i$ (or $S_{AB}^i - S_B^i$). \\
 $S^{\text{sm}}$ & Smoothed variant of an operator utilizing confidence-based gating. \\
 $S^{\text{stable}}$ & Final operator score after applying boundary stability (noise floor) logic. \\ 
\bottomrule
\end{tabular}
\end{small}
\end{table}

\subsection{\texorpdfstring{Realizing the Fusion Operator ($\cup$)}{Realizing the Fusion Operator (Union)}}

\label{subsec:fusion_operator}
The fusion operator $\cup$ instantiates a logical query $\phi$ as a single string passed to the encoder. We compare two surface realizations to assess whether NSFL is robust to the syntactic form of the fused query:
\begin{itemize}
    \item \textbf{Simple Fusion:} direct logical templates, e.g., ``$L_A$ AND $L_B$'' or ``$L_A$ AND NOT $L_B$''.
    \item \textbf{Contextual Phrasing:} natural-language formulations following the QUEST style~\citep{shen2025logicol}, e.g., ``$L_A$ that are also $L_B$'' or ``$L_A$ that are not $L_B$''.
\end{itemize}
The two realizations induce different fused scores $S_{AB}^i$ for inhibitory and disjunctive templates. NSFL yields positive gains over the monolithic baseline under both realizations, on every such template and encoder we evaluate (Appendix~\ref{app:fusion}). For pure conjunctive templates, the operator anchors on $S_M^i$, which is independent of the realization, so results are identical under both by construction.

\subsection{\texorpdfstring{Neuro-Symbolic Delta (NS-$\Delta$)}{Neuro-Symbolic Delta}}
\label{subsec:ns_delta}

The \textbf{NS-$\Delta$} quantifies the semantic contribution of an individual atom to a fused semantic representation. For a binary fusion, the score $S_{AB}^i$ admits the first-order decomposition:
\begin{equation}
    S_{AB}^i = S_A^i + \Delta_{B|A}^i = S_B^i + \Delta_{A|B}^i
\end{equation}

\paragraph{Conditional deltas.}
Beyond two atoms the conditioning context must be made explicit, since an atom's marginal semantic contribution depends on which others are already present. For an atom $X$ and a context $\mathcal{C} \subseteq \mathcal{A}$ with $X \notin \mathcal{C}$:
\begin{equation}
    \Delta_{X \mid \mathcal{C}}^i \;:=\; S_{\mathcal{C} \cup \{X\}}^i - S_{\mathcal{C}}^i
\end{equation}
where $S_{\mathcal{C}}^i$ is the similarity between item $i$ and the fusion of the atoms in $\mathcal{C}$. Setting $\mathcal{C} = \{A\}$ recovers the binary case above.

\paragraph{Rank-ordered decomposition.}
Order the atoms by ascending atomic similarity, writing $\text{minchoice}_j^i(\mathcal{A})$ for the $j$-th weakest, so that $\text{minchoice}_1^i$ is the \emph{semantic bottleneck} and $\text{minchoice}_n^i$ the dominant atom. Accumulating atoms from strongest to weakest gives contexts $\mathcal{C}_1^i = \{\text{minchoice}_n^i\}$ and $\mathcal{C}_{k+1}^i = \mathcal{C}_k^i \cup \{\text{minchoice}_{n-k}^i\}$, and the fused score telescopes:
\begin{equation}
    S_{\text{fused}}^i = S_{\mathcal{C}_1^i}^i + \sum_{k=1}^{n-1} \Delta_{\text{minchoice}_{n-k}^i \mid \mathcal{C}_k^i}^i
\end{equation}
This is an identity, not an empirical claim: it holds by construction, and reduces to Equation~(3) when $n = 2$. Its purpose is to establish that each $\Delta$ is a well-defined quantity that can be isolated and steered.

\paragraph{From decomposition to operator.}
Evaluating the chain directly would require encoding every nested subset. Our operators instead use the aggregate deficit between the fused score and the dominant atomic signal, $\mathcal{R}_n^i = S_{\text{fused}}^i - \max_j S_{A_j}^i$, which needs only scores already computed during retrieval. Since $\mathcal{R}_n^i$ pools $n-1$ marginal semantic contributions, applying it at full strength over-corrects as $n$ grows; we therefore damp it by $\gamma_n = 1/(n-1)$, recovering unit amplification for binary queries and halving it for ternary ones:
\begin{equation}
    S_{\wedge}^i = S_{\text{fused}}^i + \gamma_n \cdot \mathcal{R}_n^i, \qquad \gamma_n = \frac{1}{n-1}
\end{equation}
The value of $\gamma_n$ does not follow from the identity above; it is an operator design choice, evaluated in Section~\ref{sec:results}. This damping governs the conjunctive operators specifically; the inhibitory and disjunctive operators take structurally different forms (\S\ref{sec:operators}) and do not use $\mathcal{R}_n^i$.

\subsection{Conjectures}
\label{subsec:conjectures}

The following conjectures are testable claims about the behavior of modern contrastive encoders, motivated by the observations in Section~\ref{sec:empirical_observations}. They are working assumptions that guide the algebraic choices in Section~\ref{sec:operators}, not formal theorems. Appendix~\ref{app:principles} tests each directly against retrieval ground truth on a held-out split, independently of the operators they motivate.

\paragraph{Conjecture 1: NS-Delta Directionality (Reinforcement and Departure).} The sign of the logical delta ($\Delta_X$) serves as a latent semantic classifier. We posit that a positive delta indicates \textit{manifold reinforcement}, while a negative delta signifies \textit{semantic departure}. This builds on the "Search-with-Explanation" framework by \citet{malaviya2023quest}, which suggests that the direction of query shifts in vector space correlates with the presence or absence of grounding evidence in the retrieval corpus.

\paragraph{Conjecture 2: Asymmetric Delta-Sensitivity (The Semantic Bottleneck).} 
We posit that atomic constituents contribute unequally to fused representations. The ``weaker'' atom---the one with the lowest atomic similarity---serves as the primary constraint on logical validity, creating a \textit{semantic bottleneck} \citep{shen2025logicol, thrush2022winoground}. NSFL weights fusions by this minimal ``semantic pull'' to prevent high-confidence atoms from masking logical inconsistencies.

\paragraph{Conjecture 3: Negation as Directional Preservation.}
We posit that semantic negations in dense embeddings (e.g.\ ``not $B$'') do not induce the semantic reversal that classical Boolean logic would predict ($\neg P \equiv 1 - P$). Instead they preserve the direction of the Neuro-Symbolic Delta rather than inverting it. This absence of directional inversion is what necessitates a dedicated inhibitory operator: the encoder provides no native mechanism by which exclusion can be expressed. Appendix~\ref{app:principles} tests this directly and finds directional preservation in 69--78\% of candidates across encoders. We had additionally conjectured that negation attenuates delta magnitude; that prediction is not borne out and we do not rely on it.

Table~\ref{tab:delta_extraction} illustrates both effects on a single example.
The monolithic score for ``dog but not giraffe'' (0.26) is essentially
indistinguishable from the logically opposite ``dog and giraffe'' (0.27):
negation does not reverse the encoder's judgement, which is the behaviour
\textbf{Conjecture 3} describes and which motivates a dedicated inhibitory
operator. The unequal deltas for ``dog'' ($+0.14$) and ``man'' ($+0.09$) in
the congruent case illustrate \textbf{Conjecture 2}, showing that atomic
constituents exert differing semantic pull even when both are correctly
grounded.

\begin{table}[ht]
\centering
\caption{Empirical $\Delta$-Extraction and Non-Reversal Under Negation (BLIP-Base).}
\label{tab:delta_extraction}
\begin{small}
\begin{tabular}{@{}clllcl@{}}
\toprule
\textbf{Context} & \textbf{Atom ($L_x$)} & \textbf{$S_x$} & \textbf{Query ($L_{AB} / L_{Mono}$)} & \textbf{$S_{x}$} & \textbf{NS-$\Delta_x$} \\ \midrule
\multirow{5}{*}{\includegraphics[width=3.2cm, height=1.8cm, keepaspectratio]{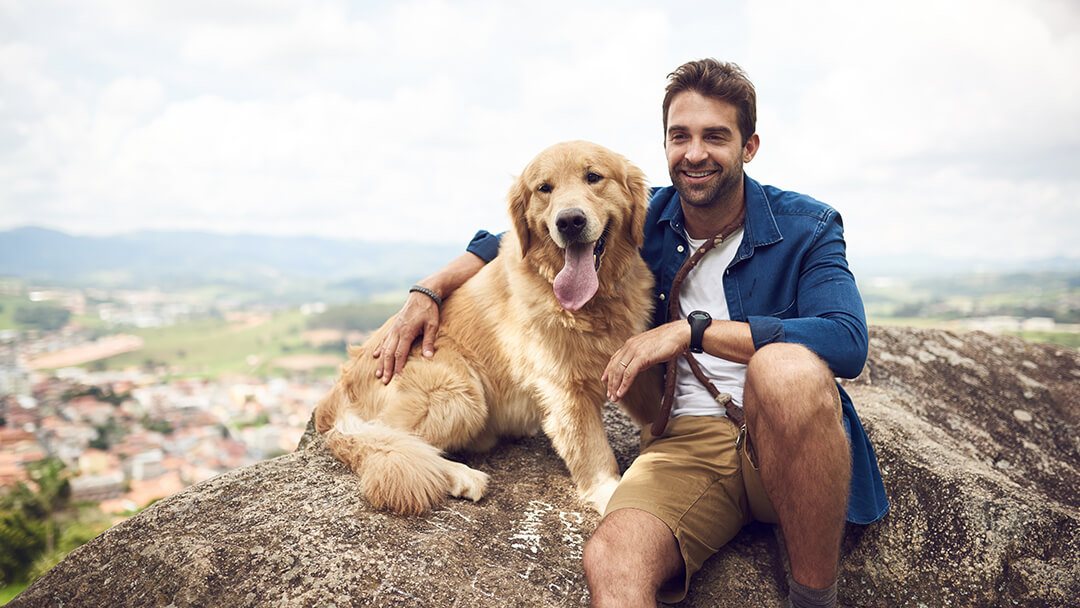}}
& dog     & 0.35 & \multirow{2}{*}{dog and giraffe} & \multirow{2}{*}{0.27} & $+0.07$ (\textbf{Pos.}) \\
& giraffe & 0.20 &                                   &                       & $-0.08$ (\textbf{Neg.}) \\ 
&         &      & dog \textbf{but not} giraffe     & 0.26                  & similar \\
\cmidrule{2-6}
& dog     & 0.35 & \multirow{2}{*}{dog and man}     & \multirow{2}{*}{0.44} & $+0.14$ (\textbf{Pos.}) \\
& man     & 0.30 &                                   &                       & $+0.09$ (\textbf{Pos.}) \\ \bottomrule
\end{tabular}
\end{small}
\end{table}


\section{Formalizing Logic for Latent Spaces}
\label{sec:formal_logic}

While logical operators have been formalized for symbolic and probabilistic domains, no comparable treatment exists for high-dimensional neural manifolds. We propose that the logical score $S_L$ for a data item $i$ can be estimated by a \textbf{functional expansion in $m$-order Neuro-Symbolic differences}:
\begin{equation}
    \label{eq:expansion}
    S_L^i = f(S_n^i) + \sum_{k=1}^m g_k\left( \Delta^{(k), i}_n \right) + \mathcal{O}\left( \Delta^{(m+1), i}_n \right)
\end{equation}
Here $n$ indexes the dominant atom ($\text{minchoice}_n$, \S\ref{subsec:ns_delta}), $f(S_n^i)$ is its zero-order similarity, $g_k$ are functions of the $k$-th order Neuro-Symbolic differences, and the terminal term is the residual of the $m$-order approximation.

We present this expansion as motivation rather than a validated framework. The residual term holds by construction, and this work investigates $m = 1$ exclusively, as higher orders require encoding nested subsets. This establishes that fuzzy-logical constructs --- t-norms and t-conorms --- can be realized as functions of semantic differences rather than raw similarities.

\subsection{Zero and First-Order Semantic Signals}
\begin{figure}[htbp]
    \centering
    \includegraphics[width=0.9\textwidth]{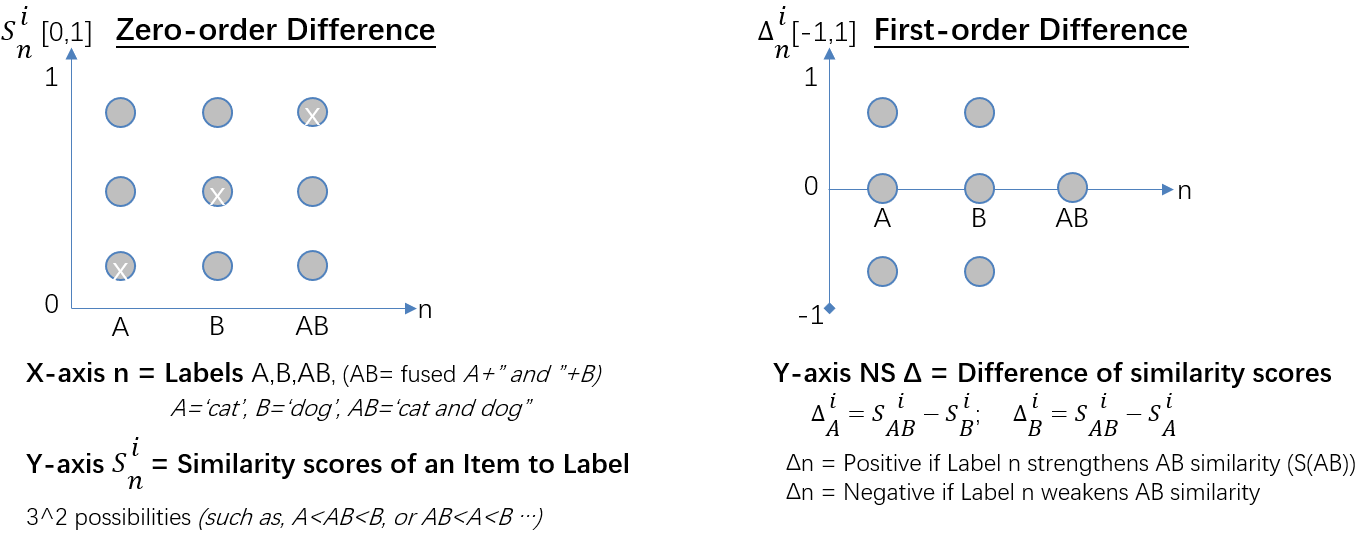}
    \caption{Comparison of zero-order similarity scores and first-order Neuro-Symbolic Deltas across retrieval candidates.}
    \label{fig:Zero-First-order-Differences-Chart}
\end{figure}

Setting $\mathcal{C} = \{A\}$ in the conditional delta of \S\ref{subsec:ns_delta} gives the first-order signal for a query with primary context $A$ and auxiliary atom $B$:
\begin{equation}
    \Delta_{B|A}^i = S_{AB}^i - S_A^i
\end{equation}
As Figure~\ref{fig:Zero-First-order-Differences-Chart} illustrates, the zero-order signal establishes an item's baseline congruence while the NS-$\Delta$ supplies a first-order correction. This lets NSFL steer a query along the manifold toward a logically congruent region rather than displacing it linearly into one the encoder does not populate.

\section{Logic Operators and Scoring Methods}
\label{sec:operators}

The \textbf{NSFL-$\Delta$ Scoring Method} is a re-ranking framework designed to maximize the score gap between items satisfying a Boolean query and those that do not. NSFL operates as a reranking layer over an oversampled candidate pool: a monolithic query retrieves $K \gg k$ candidates, each is rescored by the operator corresponding to the query's logical form, and the top $k$ are returned. Algorithm~\ref{alg:nsfl_pipeline} in Appendix~\ref{app:algorithm} gives the full procedure.

Following the functional expansion proposed in Equation~\ref{eq:expansion}, our implementation uses the first-order approximation ($m=1$), adding or subtracting NS-Deltas to zero-order similarity scores to amplify the logical margin. The general logical score $S_L$ for an item $i$ is:
\begin{equation}\label{eq:general}
    S_L(A,B) := S_{\text{base}}^L(A, B, A \cup B) + \alpha_L \cdot \delta_{\text{relevant}}^L(A,B)
\end{equation}
Where $S_{\text{base}}^L$ provides the manifold-anchored position and $\delta_{\text{relevant}}^L$ serves as the first-order corrective signal.

\subsection{Design of the Operators}
\label{subsec:operator_design}

We present the following as a family of scoring operators, each realizing a Boolean connective through a first-order correction to a fused similarity score. They are engineering realizations of fuzzy-logic principles adapted to uncalibrated vector spaces, not a formal calculus: they are not closed over $[0,1]$, they do not compose (the ternary conjunction is defined directly rather than derived from the binary one), and several deliberately sacrifice classical axioms for empirical robustness. We flag each such trade-off where it occurs and do not claim that compositions inherit t-norm guarantees.

\paragraph{Boundary stability is asymmetric by design.}
Every operator carries a stability condition, and their asymmetry reflects a divergence in how encoders handle synergistic versus inhibitory relations. For conjunction the check is \emph{upward-facing}: it protects regions of extreme reinforcement where the encoder supplies a co-occurrence reward exceeding the atomic sum, and where amplification would degrade an already-coherent representation. For negation and disjunction the checks are \emph{downward-facing}: they respond to manifold departure, enforcing a noise floor that prevents the operator from manufacturing rankings in regions where the encoder signals rejection of every constituent. Each operator below states its own condition; this asymmetry is the principle behind all of them.

\subsection{\texorpdfstring{The AND Operator ($A \land B$)}{The AND Operator (A AND B)}}
\label{op_and}
The AND operator requires finding the intersection of two concepts on the manifold. Classical fuzzy logic relies on the Gödel t-norm ($T(a,b) = \min(a,b)$), but in a deep learning context this creates a non-differentiable ridge that clips gradients and prevents smooth traversal between semantic states. Instead, NSFL leverages the encoder's native fused representation ($S_{AB}$) and evaluates its first-order difference against the dominant atomic signal, yielding a differentiable steering mechanism grounded in the parametric Hamacher t-norm and related non-compensatory operators~\citep{Klement2000}, adapted here for uncalibrated vector spaces.

\textbf{Goal:} Re-rank conjunctive queries by amplifying the marginal impact of the ``weaker'' atom (\textbf{Conjecture 2}) when the fused representation lacks synergistic reinforcement. When $S_{AB} > S_A + S_B$, the encoder's native co-occurrence reward already satisfies the conjunction; otherwise, delta-amplification ensures the semantic bottleneck drives the final score.
\begin{equation}
\begin{split}
    \Delta_{minchoice_1^i(A,B)}^i = S_{AB}^i - \max(S_A^i, S_B^i) \\
    S_{A \land B}^i = S_{AB}^i + \Delta_{minchoice_1^i(A,B)}^i = 2S_{AB}^i - \max(S_A^i, S_B^i)
\end{split}
\end{equation}

\paragraph{Boundary Stability (Noise Floor):} 
Following \textbf{Conjecture 1 (NS-Delta Directionality)}, if $S_{AB}^i > (S_A^i + S_B^i)$, we retain the inherent ``co-occurrence reward'' observed in models that exhibit bag-of-words behaviors~\citep{yuksekgonul2023when}, gracefully defaulting to the raw fused score to avoid degrading naturally cohesive concepts. Otherwise, delta-amplification enforces the logical bottleneck to prune incongruent results:
\begin{equation}
    S_{A \land B}^i = 
    \begin{cases} 
        S_{AB}^i & \text{if } S_{AB}^i > S_A^i + S_B^i \\
        2S_{AB}^i - \max(S_A^i, S_B^i) & \text{otherwise}
    \end{cases}
\end{equation}

\paragraph{Anchoring the Conjunctive Operators.}
For pure conjunctive formulas ($A \land B$ and $A \land B \land C$) we instantiate the fused score $S_{\text{fused}}^i$ with the monolithic score $S_M^i$ rather than a separately encoded fusion $S_{AB}^i$. We evaluated both and found the monolithic encoding consistently stronger, as contemporary encoders represent fluent natural-language questions more faithfully than mechanically assembled conjunctions. A consequence is that conjunctive results are \emph{invariant} to the fusion realization (\S\ref{subsec:fusion_operator}).

\paragraph{Score Range.}
The AND formula $2S_{AB}^i - \max(S_A^i, S_B^i)$ can produce values outside $[0,1]$ when $S_{AB}$ is low relative to atomic scores. We do not clip these values, as clipping would collapse distinct low-scoring candidates onto a common floor, discarding ranking information. The operators are consequently not closed over $[0,1]$, which we note as a limitation on composing them (\S\ref{sec:limitations}).

\subsection{\texorpdfstring{The NOT Operator ($A \land \neg B$)}{The NOT Operator (A AND NOT B)}}
\label{op_not}

\textbf{Goal:} Subtract the marginal contribution of $B$ from the base score of $A$, manually enforcing inhibitory behavior that dense encoders fail to induce through native negation tokens~\citep{weller2024nevir}.

\begin{equation}
    S_{A \land \neg B}^i = S_A^i - \Delta_B^i = 2S_A^i - S_{AB}^i
\end{equation}

The term $(S_{AB}^i - S_A^i)$ represents the Neuro-Symbolic Delta ($\Delta_B$)---the degree to which $B$'s presence reinforces or interferes with the signal of $A$. As posited in \textbf{Conjecture 3}, encoders preserve the direction of $\Delta_B$ under negation rather than inverting it (Appendix~\ref{app:principles}). Subtracting this delta from $S_A^i$ therefore performs the logical inversion the encoder does not, converting a directional signal into a strictly inhibitory one.

\paragraph{Smoothed Variant with Local Confidence Normalization.}
To mitigate abrupt ranking shifts, we introduce a confidence-gated variant. Because raw scores are uncalibrated (Section~\ref{sec:empirical_observations}), we define $\tilde{S}_B^i$ as a max-normalized score:
\begin{equation}
    \tilde{S}_B^i = \frac{S_B^i}{\hat{S}_B^{\max} + \epsilon}
\end{equation}
where $\hat{S}_B^{\max}$ is either the maximum $S_B$ observed in the current candidate pool, or a pre-computed corpus-level upper bound. We omit min-normalization, as filtered candidate pools exhibit selection bias that inflates the observed minimum beyond the true similarity floor.

The smoothed operator interpolates between baseline and full penalty based on local confidence:
\begin{equation}
    S_{A \land \neg B}^{\text{sm}, i} = S_A^i - \tilde{S}_B^i\left(S_{AB}^i - S_A^i\right)
\end{equation}

When $\tilde{S}_B^i \to 1$ (strongest $B$-match in pool), full negation applies; when $\tilde{S}_B^i \to 0$, the score reverts to $S_A^i$. Ablating the gate degrades both inhibitory templates slightly on BGE-Large; we retain it for stability in low-confidence regions rather than for aggregate gain. We note that the candidate pool is itself retrieved by the monolithic query and is therefore biased toward documents satisfying $A$. The pool maximum for $B$ is consequently not the corpus maximum, and $\tilde{S}_B^i$ measures confidence relative to the strongest $B$-match \emph{among candidates already plausible for $A$}. The same selection bias that motivates omitting the minimum applies in attenuated form to the maximum.

\paragraph{Boundary Stability (Noise Floor):}
Following \textbf{Conjecture 1}, when the encoder signals manifold departure ($S_{AB}^i < S_A^i$ and $S_{AB}^i < S_B^i$), we default to $S_{AB}^i$ to prevent over-rotation in regions lacking semantic grounding:
\begin{equation}
    S_{A \land \neg B}^{\text{stable}, i} = 
    \begin{cases} 
        S_{AB}^i & \text{if } S_{AB}^i < S_A^i \text{ and } S_{AB}^i < S_B^i \\
        S_A^i - \tilde{S}_B^i(S_{AB}^i - S_A^i) & \text{otherwise}
    \end{cases}
\end{equation}

\subsection{\texorpdfstring{The OR Operator ($A \lor B$)}{The OR Operator (A OR B)}}
\label{op_or}
\textbf{Goal:} Extend the Zadeh t-conorm to include fused and monolithic scores ($S_M^i$, the similarity to the full natural-language query $L_M$), allowing for the superadditivity often observed in retrieval models~\citep{zhai2023sigmoid}.
\begin{equation}
    S_{A \lor B}^i = \max(S_A^i, S_B^i, S_{AB}^i, S_M^i)
\end{equation}

Note that the fused score $S_{AB}^i$ is constructed with the same conjunctive realization used for the AND templates (\S\ref{subsec:fusion_operator}); no disjunctively phrased fusion is encoded. This is deliberate: a document satisfying both atoms necessarily satisfies $A \lor B$, so $S_{AB}^i$ enters the maximum as an additional sufficient signal, while the disjunctive constraint itself is carried by the monolithic score $S_M^i$.

\paragraph{Boundary Stability (Noise Floor):} 
Following \textbf{Conjecture 1 (NS-Delta Directionality)}, when the encoder signals a \textit{manifold departure} ($\Delta_A < 0, \Delta_B < 0$), we introduce a ``noise floor'' to respect the model's inherent rejection. This prevents the operator from ``hallucinating'' a positive result from the union of two irrelevant components.

\begin{equation}
    S_{A \lor B}^{\text{stable}, i} = 
    \begin{cases} 
        \min(S_{AB}^i, S_M^i) & \text{if } S_{AB}^i < S_A^i \text{ and } S_{AB}^i < S_B^i \\
        \max(S_A^i, S_B^i, S_{AB}^i, S_M^i) & \text{otherwise}
    \end{cases}
\end{equation}

\textbf{Note:} this boundary condition is a stability mechanism, not a t-conorm extension. It does not preserve associativity or monotonicity, and compositions of the operator do not inherit classical guarantees. The $\max$ formulation remains the operative t-conorm; the noise floor is a safeguard for the rejection regime. Since our use is single-pass reranking, where each candidate is scored once and only the induced order matters, non-associativity does not arise in practice. It would, however, obstruct any extension to nested formulas.

\subsection{\texorpdfstring{Compositional Operator AND + AND ($A \land B \land C$)}{Compositional Operator AND + AND (A AND B AND C)}}
\label{op_and_and}

\textbf{Goal:} Extend the binary conjunction to three atoms while damping the
amplification to prevent compounded over-penalization. With $n=3$ we have
$\gamma_3 = \tfrac{1}{2}$ (\S\ref{subsec:ns_delta}):
\begin{equation}
\begin{split}
    \mathcal{R}_3^i &= S_{ABC}^i - \max(S_A^i, S_B^i, S_C^i) \\
    S_{A \land B \land C}^i &= S_{ABC}^i + \tfrac{1}{2}\mathcal{R}_3^i
    = \tfrac{3}{2}S_{ABC}^i - \tfrac{1}{2}\max(S_A^i, S_B^i, S_C^i)
\end{split}
\end{equation}

\textbf{Development:} In binary conjunction the delta has one weak atom to
correct for; in ternary conjunction two potential bottlenecks are pooled into a
single deficit. Applying unit amplification to that pooled term over-penalizes
candidates whose fused representation already encodes the conjunction.

\paragraph{Boundary Stability (Noise Floor):}
Following \textbf{Conjecture 1}, we respect the synergistic alignment of the
triple-fused state when it exceeds the additive baseline:
\begin{equation}
    S_{A \land B \land C}^i =
    \begin{cases}
        S_{ABC}^i & \text{if } S_{ABC}^i > S_A^i + S_B^i + S_C^i \\
        \tfrac{3}{2}S_{ABC}^i - \tfrac{1}{2}\max(S_A^i, S_B^i, S_C^i) & \text{otherwise}
    \end{cases}
\end{equation}

\subsection{\texorpdfstring{Compositional Operator AND + AND NOT ($A \land B \land \neg C$)}{Compositional Operator AND + AND NOT (A AND B AND NOT C)}}
\textbf{Goal:} Re-rank conjunctive queries by amplifying the marginal impact of the "weaker" atoms between A and B, and as defined in \ref{op_and}, then subtract the normalized, marginal contribution of $C$:
\begin{equation}
\begin{split}
    S_{A \land B \land \neg C}^i = S_{A \land B}^i - \tilde{S}_C^i(S_{ABC}^i - S_{AB}^i) \\
\end{split}
\end{equation}

\subsection{Operators Summary}
The proposed \textbf{Neuro-Symbolic Fuzzy Logic (NSFL)} framework is summarized in Table~\ref{tab:nsfl_comprehensive}. Each operator is designed as a mathematical remedy for specific geometric failure modes identified in current dense encoders, such as \textit{semantic dilution} and \textit{negation blindness}. By grounding these operations in the \textbf{Neuro-Symbolic Delta} ($\Delta$), we ensure that the logical output remains sensitive to the marginal semantic contributions of each query constituent.

\begin{table}[ht]
\centering
\caption{Comprehensive Summary of NSFL Operators: Fuzzy Logic Formulations and Boundary Stability.}
\label{tab:nsfl_comprehensive}
\begin{small}
\begin{tabular}{@{}llll@{}}
\toprule
\textbf{Operator} & \textbf{Fuzzy Logic ($S^{\text{sm}}$)} & \textbf{Stability Trigger} & \textbf{Boundary Value ($S^{\text{stable}}$)} \\ \midrule
\textbf{AND} ($A \land B$) & $2S_{AB}^i - \max(S_A^i, S_B^i)$ & $S_{AB}^i > S_A^i + S_B^i$ & $S_{AB}^i$ (Reinforcement) \\
\textbf{NOT} ($A \land \neg B$) & $S_A^i - \tilde{S}_B^i(S_{AB}^i - S_{A}^i)$ & $S_{AB}^i < S_A^i, S_B^i$ & $S_{AB}^i$ (Departure) \\
\textbf{OR} ($A \lor B$) & $\max(S_A^i, S_B^i, S_{AB}^i, S_M^i)$ & $S_{AB}^i < S_A^i, S_B^i$ & $\min(S_{AB}^i, S_M^i)$ (Departure) \\ \bottomrule
\end{tabular}
\end{small}
\end{table}

\paragraph{Computational Overhead.}
NSFL reranking adds negligible latency: rescoring $K{=}1000$ candidates requires only 2--3 additional dot-product operations per atom, completing in microseconds on CPU. The dominant cost remains the initial ANNS retrieval.

\section{Spherical Query Optimization (SQO)}
\label{sec:sqo}

Standard ANNS indices are natively optimized for single-objective distance minimization, making them structurally incompatible with the non-linear requirements of fuzzy logical constraints. To bridge this gap, we propose \textbf{Spherical Query Optimization (SQO)}, a pre-search query transformation that translates a complex logical formula into a single optimal theoretical query vector.

\subsection{Optimization on the Unit Hypersphere}
\label{sec:sqo_optimization}

\noindent
\begin{minipage}[t]{0.62\textwidth}
    \vspace{0pt} 
    We assume a corpus $\mathcal{D}$ where item vectors are normalized to the unit hypersphere $S^{d-1}$. Given a logical formula, we define a continuous and differentiable scoring function $f(x)$ using our proposed \textbf{NSFL} operators. We then solve for the optimal theoretical solution: 
    \begin{equation}
        x^* = \argmax_{x \in S^{d-1}} f(x)
    \end{equation}

    \vspace{1em}
    This optimization is performed via \textbf{Riemannian Stochastic Gradient Descent (RSGD)}. To maintain $x \in S^{d-1}$ without numerical drift, we utilize a formal \textbf{retraction map} $R_x: T_x S^{d-1} \to S^{d-1}$ \citep{absil2009optimization}. For the unit hypersphere, this is computed as the $L_2$ normalization of the tangent update:
    \begin{equation}
        R_x(\eta) = \frac{x + \eta}{\|x + \eta\|}, \quad \text{where } \eta = \alpha \nabla_R f(x)
    \end{equation}
\end{minipage}
\hfill
\begin{minipage}[t]{0.33\textwidth}
    \vspace{0pt}
    \centering
    \includegraphics[width=\textwidth]{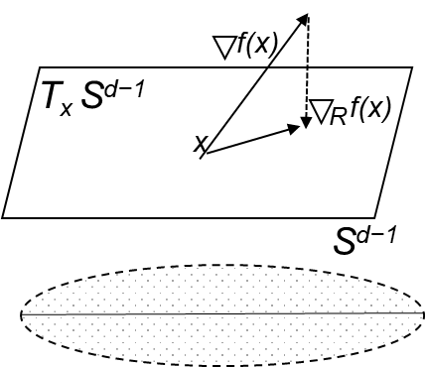}
    \captionof{figure}{Projection of Euclidean gradient $\nabla f(x)$ onto $T_x S^{d-1}$ during SQO.}
    \label{fig:rgd}
\end{minipage}

As illustrated in Figure~\ref{fig:rgd}, RSGD updates the candidate vector $x$ by projecting the Euclidean gradient $\nabla f(x)$ onto the tangent space $T_x S^{d-1}$, followed by the retraction $R_x$ back to the manifold surface. The complete implementation logic and algorithmic steps are detailed in \textbf{Algorithm~\ref{alg:sqo_logic}} (see Appendix~\ref{appendix:reproducibility}). 

Since our scoring functions $f(x)$ are compositions of $N$ atomic vectors in $d$-dimensional space, each RSGD iteration maintains a linear computational complexity of $O(N \cdot d)$ \citep{bonnabel2013stochastic}. In practice, convergence is rapid: empirical seed-variance analysis confirms mean per-query runtime under $1$ms on CPU with negligible score variance across initializations ($<0.1\%$). Hyperparameter configurations are provided in Table~\ref{tab:sqo_hyperparams}.

\subsection{Hybrid Retrieval Strategy}
Because our fuzzy logic scoring functions are continuous, vectors proximal to $x^*$ are highly likely to be strong logical matches. We employ a two-step hybrid process to bridge the gap between theoretical optima and practical index constraints: (1) ANNS search using $x^*$ to retrieve top $K$ neighbors, and (2) local rescoring of these $K$ candidates using the full $f(x)$ formulation to identify the best match. SQO allows legacy ANNS indices to support complex logical constraints without requiring structural modification or expensive data-dependent pre-processing.

\section{Experimental Setup}
\label{sec:experiments}

To evaluate the generalizability of the \textbf{NSFL} framework, we designed a cross-modal benchmarking suite covering text-to-text and text-to-image retrieval.

\paragraph{Datasets:} 
1) \textbf{QUEST}~\citep{malaviya2023quest}: A text-to-text benchmark using a 325k Wikipedia corpus. We utilize the 323 validation queries for modular ablation and the \textbf{1,727 test queries} for primary evaluation across six logical templates. 
2) \textbf{COCO-LOGIC}: A text-to-image benchmark derived from the \textbf{COCO-2017} validation set (5k images)~\citep{lin2014microsoft}. We generated 600 queries across six templates (100 each) to evaluate visio-linguistic compositionality. Ground truth is strictly mapped to COCO multi-label annotations; for instance, $A \land \neg B$ is satisfied only if category $A$ is present and $B$ is absent. COCO-LOGIC (queries and ground-truth mappings) is released with this paper at \url{https://github.com/vladiaidev-lab/NSFL}; the COCO-2017 images themselves are obtained from the original source.

We focus on QUEST and COCO-LOGIC as they provide multi-template logical queries with explicit ground truth. We additionally evaluate on NevIR~\citep{weller2024nevir} (Appendix~\ref{app:nevir}), a pairwise negation benchmark, where NSFL's inhibitory operator yields consistent improvements of \textbf{+7\% to +10.5\%} in mean per-query accuracy and \textbf{+19\% to +25\%} under NevIR’s stricter pairwise criterion, across encoders.

\paragraph{Parameters:}
NSFL introduces no parameters fitted to evaluation data. Three constants appear in the operators and we state them explicitly. The amplification coefficient $\alpha_L$ in Equation~\ref{eq:general} is fixed to unity for all operators. The damping factor is $\gamma_n = 1/(n-1)$, determined by atom count rather than fitted, giving $\gamma_2 = 1$ and $\gamma_3 = 1/2$. The numerical stabiliser in the confidence gate is $\epsilon = 10^{-8}$; results are insensitive to it over several orders of magnitude. The single choice made using held-out data is the fusion realization, selected on the QUEST validation split and applied globally (Appendix~\ref{app:fusion}).

\paragraph{Zero-Shot Evaluation Protocol:} 
The stability triggers and inhibitory logic derived in Section \ref{sec:operators} are applied directly to raw encoder outputs without fitting parameters to the evaluation data. We evaluate this ``zero-shot'' logical correction against a standard neural baseline using two pipelines:
\begin{itemize}
    \item \textbf{Baseline Retrieval:} Performs top-$k$ retrieval ($k=100$) using \textit{Monolithic Query} ($L_M$) score $S_M^i$.
    \item \textbf{NSFL Reranking:} Performs an initial ``over-sampling'' ($K=1000$) based on $L_M$, followed by \textbf{NSFL} rescoring using atomic ($\Delta$) and stability triggers. Final top-$k$ ($k=100$) items are re-selected from this corrected pool.
\end{itemize}

\textbf{Evaluation Metric:} We utilize \textit{Mean Average Precision} (mAP) as our primary metric. Unlike Recall@$k$, mAP is sensitive to the precise rank-order, capturing the model's ability to prioritize semantically grounded results over logical hallucinations.

\subsection{Baselines for Comparison}
\label{subsec:baselines}

To isolate the specific contributions of the \textbf{NSFL} framework, we compare our results against three distinct baseline strategies, ranging from standard neural outputs to naive linear geometric modifications.

\paragraph{1. Monolithic Baseline (Standard):} 
The primary point of comparison is the raw output of the encoder using the fused or natural language query ($L_M$) without any post-hoc intervention. This represents the current state-of-the-art approach in neural retrieval, where logical constraints are handled implicitly by the model's pre-training on complex phrases.

\paragraph{2. Linear Geometric Baselines:} 
We evaluate two linear strategies to determine if logical intent can be recovered through simple vector arithmetic on the unit hypersphere, a common heuristic in distributional semantics \citep{mitchell2008vector, kanerva2009hyperdimensional}:

\begin{itemize}
    \item \textbf{Orthogonal Projection (Gram-Schmidt):} For inhibitory constraints (e.g., $L_A \text{ NOT } L_B$), we apply Gram-Schmidt orthogonalization \citep{strang2006} to maximize the representation of the primary atom $\mathbf{v}_A$ by removing its projection onto the negated atom $\mathbf{v}_B$:
    \begin{equation}
        S_{A,B} = \mathbf{v}_A^\top \mathbf{v}_B, \quad \mathbf{v}_{temp} = \mathbf{v}_A - S_{A,B} \mathbf{v}_B, \quad \mathbf{v}^* = \frac{\mathbf{v}_{temp}}{\|\mathbf{v}_{temp}\|_2}
    \end{equation}
    
    \item \textbf{Normalized Vector Addition:} For conjunctive constraints (e.g., $L_A \text{ AND } L_B$), we evaluate the linear summation of atomic embeddings followed by re-normalization \citep{mitchell2008vector}:
    \begin{equation}
        \mathbf{v}_{temp} = \mathbf{v}_A + \mathbf{v}_B, \quad \mathbf{v}^* = \frac{\mathbf{v}_{temp}}{\|\mathbf{v}_{temp}\|_2}
    \end{equation}
\end{itemize}

\paragraph{Why Not Set Intersection or Sparse Systems?}
Set intersection is ill-suited to dense retrieval because continuous scores lack a natural membership threshold, and intersection discards the relative ranking information that NSFL exploits. Sparse retrieval (BM25, SPLADE) and late-interaction models (ColBERT) handle term matching well but require separate indices and infrastructure. NSFL targets a different deployment scenario: augmenting existing dense retrieval pipelines without retraining or index restructuring. We position NSFL as complementary to these alternatives, offering a lightweight post-hoc solution for systems already committed to dense single-vector retrieval.

\subsection{Correction Tiers and Ablations}
\label{subsec:ablation_tiers}
Our protocol distinguishes between two operational stages to isolate the impact of logical intervention:
\begin{enumerate}
    \item \textbf{Pre-ANNS Query Optimization:} Evaluates query-side adjustments—specifically \textit{NSFL Query Optimization on Sphere} (SQO)—applied prior to the initial vector search.
    \item \textbf{Post-ANNS Reranking:} Our primary intervention, applying fuzzy logic operators to the top $K=1000$ candidates. We compare this against \textit{Geometric Baselines} in our ablation studies (Table~\ref{tab:matrix_ablation}) to evaluate the necessity of non-linear intervention.
\end{enumerate}

\subsection{Results}
\label{sec:results}

\begin{table}[H]
\centering
\caption{Mean Average Precision (mAP@100) across logical templates. We compare the \textit{Monolithic Baseline} (raw encoder) against \textit{NSFL Reranking}. \textbf{Rel. $\Delta$} indicates the percentage improvement over the baseline.}
\label{tab:main_results}
\begin{adjustbox}{width=\textwidth}
\begin{small}

\begin{tabular}{@{}llcccccc|c@{}}
\toprule
\textbf{Dataset} & \textbf{Model} & $A \land B$ & $A \land B \land C$ & $A \land \neg B$ & $A \land B \land \neg C$ & $A \lor B$ & $A \lor B \lor C$ & \textbf{Avg.} \\ \midrule
\multicolumn{9}{l}{\textit{\textbf{Text-to-Text (QUEST)}}} \\ \midrule
\texttt{BGE-Large-v1.5} & Baseline & 0.0384 & 0.0423 & 0.0501 & 0.0153 & 0.1352 & 0.1075 & 0.0648 \\
 & NSFL (Ours) & \textbf{0.0525} & \textbf{0.0445} & \textbf{0.0753} & \textbf{0.0247} & \textbf{0.1440} & \textbf{0.1348} & \textbf{0.0793} \\
 & \textit{Rel. $\Delta$} & \textit{+37\%} & \textit{+5\%} & \textit{+50\%} & \textit{+61\%} & \textit{+7\%} & \textit{+25\%} & \textit{+22\%} \\ \midrule

\texttt{Nomic-Embed-v2} & Baseline & 0.0325 & 0.0321 & 0.0536 & 0.0187 & 0.1578 & 0.1258 & 0.0701 \\
 & NSFL (Ours) & \textbf{0.0379} & \textbf{0.0356} & \textbf{0.0929} & \textbf{0.0338} & \textbf{0.1602} & \textbf{0.1408} & \textbf{0.0835} \\
 & \textit{Rel. $\Delta$} & \textit{+16\%} & \textit{+11\%} & \textit{+73\%} & \textit{+81\%} & \textit{+1\%} & \textit{+12\%} & \textit{+19\%} \\ \midrule

\texttt{E5-base-v2} & Baseline & 0.0450 & 0.0475 & 0.0569 & 0.0202 & 0.1407 & 0.1039 & 0.0690 \\
 & NSFL (Ours) & \textbf{0.0522} & \textbf{0.0479} & \textbf{0.0801} & \textbf{0.0310} & \textbf{0.1495} & \textbf{0.1407} & \textbf{0.0836} \\
 & \textit{Rel. $\Delta$} & \textit{+16\%} & \textit{+1\%} & \textit{+41\%} & \textit{+54\%} & \textit{+6\%} & \textit{+35\%} & \textit{+21\%} \\ \midrule

\texttt{LogiCoL-E5-v2} & Baseline & 0.0827 & \textbf{0.0946} & 0.1135 & 0.0519 & 0.2147 & 0.1534 & 0.1185 \\
 & NSFL (Ours) & \textbf{0.0850} & 0.0931 & \textbf{0.1346} & \textbf{0.0762} & \textbf{0.2491} & \textbf{0.2102} & \textbf{0.1414} \\
 & \textit{Rel. $\Delta$} & \textit{+3\%} & \textit{-2\%} & \textit{+19\%} & \textit{+47\%} & \textit{+16\%} & \textit{+37\%} & \textit{+19\%} \\ \midrule

\multicolumn{9}{l}{\textit{\textbf{Image-to-Text (COCO)}}} \\ \midrule
\texttt{BLIP-Large} & Baseline & 0.1420 & 0.1071 & 0.3227 & 0.1542 & 0.6221 & 0.6221 & 0.3284 \\
 & NSFL (Ours) & \textbf{0.1795} & \textbf{0.1272} & \textbf{0.5141} & \textbf{0.2620} & \textbf{0.7297} & \textbf{0.8181} & \textbf{0.4384} \\
 & \textit{Rel. $\Delta$} & \textit{+26\%} & \textit{+19\%} & \textit{+59\%} & \textit{+70\%} & \textit{+17\%} & \textit{+31\%} & \textit{+34\%} \\ \midrule
\texttt{SigLIP} & Baseline & \textbf{0.1663} & \textbf{0.0928} & 0.2818 & 0.1370 & 0.7345 & 0.7646 & 0.3628 \\
 & NSFL (Ours) & 0.1548 & 0.0857 & \textbf{0.4608} & \textbf{0.1670} & \textbf{0.7535} & \textbf{0.8258} & \textbf{0.4079} \\
 & \textit{Rel. $\Delta$} & \textit{-7\%} & \textit{-8\%} & \textit{+64\%} & \textit{+22\%} & \textit{+3\%} & \textit{+8\%} & \textit{+12\%} \\ 
\bottomrule
\end{tabular}
\end{small}
\end{adjustbox}
\end{table}

\textit{Note: All NSFL results use the Contextual Phrasing realization of the fusion operator (\S\ref{subsec:fusion_operator}), selected by pooled validation-set performance across encoders and applied uniformly to every encoder and template. Appendix~\ref{app:fusion} reports both realizations separately, the selection procedure, and the cost of this policy relative to a per-cell oracle.}

Improvements are statistically significant (Wilcoxon signed-rank, $p < 0.01$) for the inhibitory and ternary disjunctive templates across all encoders. Gains on binary conjunction and binary disjunction are smaller and reach significance in some configurations but not others. Appendix~\ref{app:significance} reports every configuration individually, together with bootstrap confidence intervals. Gains are strongest and most consistent on inhibitory templates, a pattern that also holds on the NevIR negation benchmark.

\FloatBarrier

\paragraph{Comparison with Classical Score-Combination Baselines.}
To isolate the contribution of the Neuro-Symbolic Delta specifically, we compare against classical fuzzy operators applied directly to atomic similarity scores. These baselines rerank the identical candidate pool under the identical evaluation protocol and differ only in that they have no access to the fused representation. The gains are likewise not an artifact of oversampling: reranking the baseline's own top-100 pool recovers 95.7\% of $K{=}1000$ performance on LogiCoL-E5-v2 and 98.0\% on BGE-Large (Appendix~\ref{app:k_ablation}). We include a calibrated variant, in which each atom's scores are min--max normalized within the candidate pool before the operator is applied, so that the comparison does not rest on the uncalibrated scale of raw similarities.

\begin{table}[!ht]
\centering
\caption{Classical score-combination baselines versus NSFL on QUEST (mAP@100, averaged over the six logical templates). All methods rerank the same $K{=}1000$ candidate pool.}
\label{tab:tnorm_baselines}
\begin{small}
\begin{tabular}{@{}llccc@{}}
\toprule
\textbf{Method} & \textbf{Uses $S_{\text{fused}}$} & \textbf{BGE-Large} & \textbf{E5-base-v2} & \textbf{LogiCoL-E5-v2} \\ \midrule
Monolithic baseline & --- & 0.0648 & 0.0690 & 0.1185 \\ \midrule
Gödel ($\min$/$\max$) & No & 0.0555 & 0.0594 & 0.0977 \\
Product t-norm & No & 0.0535 & 0.0543 & 0.1059 \\
Łukasiewicz & No & 0.0626 & 0.0608 & 0.1107 \\
Gödel, pool-normalized & No & 0.0555 & 0.0618 & 0.0987 \\ \midrule
\textbf{NSFL (ours)} & Yes & \textbf{0.0793} & \textbf{0.0836} & \textbf{0.1414} \\
\bottomrule
\end{tabular}
\end{small}
\end{table}

Two observations follow. First, NSFL outperforms every classical combination including the calibrated variant --- on BGE-Large by 43\% relative over pool-normalized Gödel --- indicating that the fused-score signal contributes beyond what calibration alone provides. Second, and more strikingly, several classical baselines score \emph{below} the monolithic baseline: naive fuzzy combination over atomic scores can be worse than ignoring logical structure entirely. The gains reported in Table~\ref{tab:main_results} are therefore not attributable to logical decomposition as such, but to the specific operators we define. The Łukasiewicz family additionally illustrates the calibration problem directly: because $\max(0, S_A + S_B - 1)$ requires similarities approaching unity, it collapses to zero for 22--41\% of candidates depending on encoder, destroying the ranking.

\paragraph{Additive Gains on SOTA.}
Applying NSFL to \texttt{LogiCoL-E5-v2} yields a further \textbf{+19\% average improvement} over the already-enhanced baseline (0.1185 $\to$ 0.1414 mAP@100), with gains up to \textbf{+47\%} on $A \land B \land \neg C$. This confirms NSFL is complementary to training-time optimizations.

\paragraph{On Absolute Magnitudes.}
Absolute mAP on QUEST is low for every method we evaluate, including the monolithic baseline and the logic-fine-tuned LogiCoL encoder. This reflects the difficulty of the benchmark: queries target set-valued answers over a 325k-document corpus, with few relevant documents per query. We do not claim NSFL makes these queries easy. Relative improvement on a low base can mislead, so we report three complementary views. Recall@$K$ (Appendix~\ref{sec:recall_comparison}) confirms the same ordering under a metric directly comparable to prior work. Per-query significance testing (Appendix~\ref{app:significance}) establishes which gains are systematic rather than aggregate artifacts. And on NevIR (Appendix~\ref{app:nevir}), where absolute accuracies are high and directly interpretable, NSFL improves pairwise negation accuracy by \textbf{+7\% to +10.5\%} absolute across three encoders --- on BGE-Large from 0.611 to 0.717 --- and by \textbf{+19\% to +25\%} under the stricter criterion that both documents in a pair rank correctly, from 0.256 to 0.508. We regard the NevIR result as the clearest demonstration that the inhibitory operator recovers signal the encoder possesses but does not act on.

\paragraph{Encoder-Specific Behavior.}
\texttt{SigLIP}'s sigmoid objective causes regression on binary conjunction ($-7\%$), but inhibitory templates recover strongly ($A \land B \land \neg C$: $+22\%$), demonstrating NS-Delta steering remains effective even with suppressed base signals.

\paragraph{Modular Tiers.}
Table~\ref{tab:matrix_ablation} shows NSFL outperforming the geometric baseline in every configuration, by 0.0113 mAP@100 on QUEST and 0.0204 on COCO-LOGIC under reranking. Geometric orthogonalization recovers part of the inhibitory signal but degrades conjunctive templates (Appendix~\ref{app:ablation_detail}), supporting our use of score-level operators over vector-space manipulation.

\begin{table}[ht]
\centering
\caption{Systematic ablation of modular tiers: mAP@100 averaged over the six
logical query types, comparing Pre-ANNS Optimization and Post-ANNS Reranking
on BGE/QUEST (text) and BLIP-Large/COCO-Logic (cross-modal). Best non-baseline
value per dataset in \textbf{bold}. See Table~\ref{tab:ablation_full} in
Appendix~\ref{app:ablation_detail} for the per-logic-type breakdown.}
\label{tab:matrix_ablation}
\begin{small}
\begin{tabular}{@{}llcc cc@{}}
\toprule
& & \multicolumn{2}{c}{\textbf{Configuration}} & \multicolumn{2}{c}{\textbf{Avg. mAP@100}} \\
\cmidrule(lr){3-4} \cmidrule(lr){5-6}
\textbf{Method} & \textbf{Variant} & \textbf{Pre-Opt.} & \textbf{Post-Rerank} & \textbf{QUEST} & \textbf{COCO-L} \\ \midrule
Baseline & Monolithic & --- & --- & 0.0648 & 0.3284 \\ \midrule
\multirow{3}{*}{\textbf{NSFL}}
 & Rerank only & ---        & \checkmark & \textbf{0.0793} & 0.4384 \\
 & Opt.\ only  & \checkmark & ---        & 0.0696 & 0.3790 \\
 & Hybrid      & \checkmark & \checkmark & 0.0792 & \textbf{0.4387} \\ \midrule
\multirow{3}{*}{\textbf{Geometric}}
 & Rerank only & ---        & \checkmark & 0.0680 & 0.4180 \\
 & Opt.\ only  & \checkmark & ---        & 0.0594 & 0.3660 \\
 & Hybrid      & \checkmark & \checkmark & 0.0654 & 0.4165 \\ \bottomrule
\end{tabular}
\end{small}
\end{table}

\paragraph{Spherical Query Optimization.}
Table~\ref{tab:matrix_ablation} separates the two operational stages. Applied alone, SQO improves over the monolithic baseline --- from 0.0648 to 0.0696 mAP@100 on QUEST and from 0.3284 to 0.3790 on COCO-LOGIC --- at the cost of a single ANNS query and no post-retrieval computation. Applied together with reranking it yields no further gain: the Hybrid configuration matches Rerank-only on every template whose objective SQO represents exactly. We take this as a substantive rather than a null finding. SQO recovers a meaningful fraction of NSFL's improvement at standard retrieval latency, which is the regime where reranking $K{=}1000$ candidates may be infeasible; but where reranking is affordable it subsumes what SQO provides. The two are alternatives along a latency--accuracy trade-off rather than complementary stages.

\paragraph{What a single query vector can and cannot express.}
The correspondence is exact only for the pool-independent operators. The inhibitory operators depend on a confidence gate normalized against the candidate pool (\S\ref{op_not}), and that quantity does not exist before retrieval: computing it would require the very monolithic search SQO is intended to replace. SQO therefore optimizes the ungated form of those two operators, and they are precisely the templates on which Hybrid and Rerank-only diverge, by $+0.0009$ and $-0.0015$ mAP@100 respectively. We regard this as a principled boundary on single-vector approximation rather than an implementation limitation: operator components defined over the retrieved neighbourhood are not expressible in a pre-retrieval objective.

\FloatBarrier

\section{Limitations and Broader Impacts}
\label{sec:limitations}

\paragraph{The Conjunction Gap.}
Conjunctive operators remain the weakest category. On QUEST, $A \land B$ and $A \land B \land C$ reach only 0.036--0.093 mAP@100, against 0.075--0.135 for inhibitory and 0.135--0.249 for disjunctive templates. While conjunction is intrinsically strict (admitting no partial match), per-query analysis reveals a specific failure mechanism: losses are highly non-uniform. Severe degradations occur when the candidate pool is densely clustered and relevant documents are sparse, causing a tiny score perturbation to displace a relevant document by several ranks and collapse its Average Precision (Appendix~\ref{app:significance}).

\paragraph{Toward Significance-Gated Correction.}
This failure profile suggests a remedy. NSFL applies its correction wherever the deficit $\mathcal{R}_n^i$ is nonzero, even when it is indistinguishable from local noise. We tested scaling the correction by its significance relative to the pool dispersion $\sigma_{\mathcal{K}}$:
\begin{equation}
    S_{\wedge}^i = S_{\text{fused}}^i + \gamma_n \cdot \frac{|\mathcal{R}_n^i|}{|\mathcal{R}_n^i| + \sigma_{\mathcal{K}}} \cdot \mathcal{R}_n^i
\end{equation}
This variant does not improve aggregate mAP, but behaves as designed: among the five largest per-query losses on $A \land B \land C$, three shrink by 33--40\%. The worst case is unrescued because its deficit is large relative to $\sigma_{\mathcal{K}}$---a substantive signal traversing a dense region rather than noise. This confirms that a component of the gap stems from unconditional correction in low-signal regions, though closing the remainder requires future work (e.g., encoder-adaptive damping or distributional gates).

\paragraph{Structural Limitations.}
The operators are not closed over $[0,1]$ (\S\ref{op_and}) and are defined per arity rather than compositionally; applying the binary operator twice for ternary conjunction degrades performance. Combined with the axiom sacrifices in \S\ref{op_or}, NSFL covers a fixed set of logical templates rather than arbitrary nested formulas.

\paragraph{Ethical Considerations and Bias Amplification.}
Because NSFL inherits biases from the encoder, the \texttt{NOT} operator poses a specific risk: excluding concept $B$ may disproportionately suppress a protected demographic spuriously correlated with $B$. We recommend deployment be accompanied by bias audits measuring retrieval rates for protected-attribute proxies before and after applying inhibitory operators.

\section{Conclusion}
\label{sec:conclusion}

We introduced \textbf{Neuro-Symbolic Fuzzy Logic (NSFL)}, a training-free first-order fuzzy-logical framework for frozen dense retrievers. Its foundation is the \textbf{Neuro-Symbolic Delta}, an exact first-order decomposition of a fused similarity score into its atoms' contributions. On that foundation we build operators for conjunction, inhibition, and disjunction. Disjunction extends the Zadeh t-conorm; conjunction inherits the non-compensatory principle that distinguishes t-norms from averaging operators, though it is not itself a t-norm; and inhibition departs from fuzzy negation deliberately, since encoders do not implement complement (\S\ref{sec:empirical_observations}). The shift from zero-order similarity to first-order differences is what makes non-convex steering possible: inhibition in particular is structurally inaccessible to zero-order scoring, however it is combined.

Across six encoders and two modalities, gains are largest on inhibitory and disjunctive templates, reaching \textbf{+81\%} relative improvement, and hold additively on an encoder already fine-tuned for logical retrieval. On NevIR, where absolute accuracies are interpretable, NSFL raises pairwise accuracy, NevIR’s both-correct criterion, from \textbf{0.256 to 0.508}. The decomposition is exact and general; the operators built on it are engineering realizations rather than a closed algebra --- not closed over $[0,1]$, defined per arity rather than derived compositionally, and trading classical axioms for robustness at their boundary conditions.

\textbf{Spherical Query Optimization (SQO)} projects a logical formula into a single query vector, improving over monolithic retrieval at standard ANNS latency; reranking subsumes its gains where affordable, making the two alternatives along a latency--accuracy trade-off. Operator components that depend on statistics of the retrieved neighborhood are not expressible in a pre-retrieval objective at all, which delimits single-vector approximation.

Finally, we locate the mechanism behind the ``Conjunction Gap'': corrections apply unconditionally, including where the deficit is indistinguishable from noise and a few ranks of displacement are costly. A significance-gated variant mitigates the largest failures without improving aggregate accuracy, indicating more than one cause. First-order analysis of fused representations recovers substantially more logical structure than zero-order similarity affords; the arity and composition limits we document mark where a training-free treatment currently stops.

\section*{Acknowledgments}
We thank Dr.\ Joseph Kampeas for developing the geometric methods that serve as a foundational baseline in this work. The authors thank Lei Zhou for scientific background synthesis and internal technical notes that shaped the direction of this work, and for his engagement with the project throughout. We thank Dr.\ Efthymia Tsamoura for valuable scientific discussions and neuro-symbolic background insights. We thank Shir Niego Komforti and Aviad Cohen Zada for their contributions to the preparation of the COCO-LOGIC dataset, and Ilan Bronshtein for the experimental integration into the GaussDB-Vector database \citep{GaussDB-Vector-2025}. We also thank Dr.\ Eliezer Levy, CTO and director of the Pnueli Lab, for his continuous support and encouragement throughout this project.

\bibliography{main}
\bibliographystyle{tmlr}

\appendix
\section*{Appendix}

\section{The NSFL Reranking Pipeline}
\label{app:algorithm}

Algorithm~\ref{alg:nsfl_pipeline} states the procedure as implemented. Two features are worth noting because they are not evident from the operator definitions alone. First, the confidence gate used by the inhibitory operators requires a per-query pool statistic, computed after retrieval and before rescoring; this is what makes those operators inexpressible in a pre-retrieval objective (\S\ref{sec:sqo}). Second, the operators do not share a common functional form: conjunction applies an additive damped correction, disjunction takes a maximum over available signals, and inhibition applies a gated subtraction. We therefore state each explicitly rather than presenting a single parameterized rule.

\begin{algorithm}[H]
\caption{NSFL Reranking Pipeline}
\label{alg:nsfl_pipeline}
\begin{algorithmic}[1]
\Require Formula $\phi$ over atoms $\{A_1,\dots,A_n\}$; index $\mathcal{I}$;
         pool size $K$; output size $k$ ($K \gg k$)
\Ensure Top-$k$ items ranked by $S_\phi$ 
\Statex \textbf{Stage 1: Query realization}
\State $\mathbf{v}_M \gets \textsc{Encode}(\text{question text})$ \State $\mathbf{v}_{A_j} \gets \textsc{Encode}(A_j)$ for $j = 1,\dots,n$ \If{$n = 2$}
    \State $\mathbf{v}_{AB} \gets \textsc{Encode}(\textsc{Fuse}(A_1, A_2))$ \Else
    \State $\mathbf{v}_{AB}, \mathbf{v}_{AC}, \mathbf{v}_{BC} \gets$            \textsc{Encode} of each pairwise fusion
    \State $\mathbf{v}_{ABC} \gets \textsc{Encode}(\textsc{Fuse}(A_1, A_2, A_3))$ \EndIf
\Comment{\textsc{Fuse} is conjunctive for every $\phi$, \S\ref{subsec:fusion_operator}}

\Statex \textbf{Stage 2: Candidate generation}
\State $\mathbf{q} \gets \mathbf{v}_M$        \Comment{or $x^*$ from SQO, \S\ref{sec:sqo}}
\State $\mathcal{K}, \{S_M^i\} \gets \textsc{ANNS}(\mathcal{I}, \mathbf{q}, K)$ 
\Statex \textbf{Stage 3: Score extraction}
\For{each $i \in \mathcal{K}$}
    \State $S_{A_j}^i \gets \langle \mathbf{v}_{A_j}, \mathbf{v}_i \rangle$ for all $j$     \State $S_{AB}^i$, and for $n{=}3$ also $S_{AC}^i, S_{BC}^i, S_{ABC}^i$ \EndFor

\Statex \textbf{Stage 4: Pool statistics}
\For{$j = 1,\dots,n$}
    \State $\hat{S}_{A_j} \gets \max_{i \in \mathcal{K}} S_{A_j}^i$     \State $\tilde{S}_{A_j}^i \gets S_{A_j}^i / (\hat{S}_{A_j} + \epsilon)$            \Comment{confidence gate; requires the pool}
\EndFor

\Statex \textbf{Stage 5: Rescoring}
\For{each $i \in \mathcal{K}$}
    \If{$\phi = A \land B$}
        \State $S_\phi^i \gets 2S_M^i - \max(S_A^i, S_B^i)$;
               \textbf{if} $S_M^i > S_A^i + S_B^i$ \textbf{then} $S_\phi^i \gets S_M^i$     \ElsIf{$\phi = A \land B \land C$}
        \State $S_\phi^i \gets S_M^i + \tfrac{1}{2}(S_M^i - \max_j S_{A_j}^i)$;
               \textbf{if} $S_M^i > \sum_j S_{A_j}^i$ \textbf{then} $S_\phi^i \gets S_M^i$     \ElsIf{$\phi = A \land \neg B$}
        \State $S_\phi^i \gets S_A^i - \tilde{S}_B^i(S_{AB}^i - S_A^i)$;
               \textbf{if} $S_{AB}^i < S_A^i, S_B^i$ \textbf{then} $S_\phi^i \gets S_{AB}^i$     \ElsIf{$\phi = A \land B \land \neg C$}
        \State $t \gets 2S_{AB}^i - \max(S_A^i, S_B^i)$;
               \textbf{if} $S_{AB}^i > S_A^i + S_B^i$ \textbf{then} $t \gets S_{AB}^i$         \State $S_\phi^i \gets t - \tilde{S}_C^i(S_{ABC}^i - S_{AB}^i)$     \ElsIf{$\phi = A \lor B$}
        \State $S_\phi^i \gets \max(S_M^i, S_{AB}^i, S_A^i, S_B^i)$;
               \textbf{if} $S_{AB}^i < S_A^i, S_B^i$ \textbf{then} $S_\phi^i \gets S_{AB}^i$     \ElsIf{$\phi = A \lor B \lor C$}
        \State $S_\phi^i \gets \max(S_M^i, S_{ABC}^i, S_{AB}^i, S_{AC}^i, S_{BC}^i, S_A^i, S_B^i, S_C^i)$     \EndIf
\EndFor

\Statex \textbf{Stage 6: Selection}
\State \Return top-$k$ of $\mathcal{K}$ by descending $S_\phi^i$ \end{algorithmic}
\end{algorithm}

\paragraph{Conjunctive anchoring.}
The conjunctive operators use $S_M^i$ where the operator definitions in \S\ref{sec:operators} write $S_{\text{fused}}^i$. The separately encoded $S_{AB}^i$ and $S_{ABC}^i$ are computed but unused for these two templates. As noted in \S\ref{sec:operators}, we evaluated both and retained the monolithic anchoring; this is also why conjunctive results are invariant to the fusion realization.

\section{Geometric Justification of Asymmetric Delta-Sensitivity}
\label{app:geometric_justification}

To formalize the intuition behind Conjecture 2, we consider the neural retrieval process as a projection within a high-dimensional Hilbert space $\mathcal{H} \cong \mathbb{R}^d$. We represent atomic query constituents as unit vectors $\mathbf{a}, \mathbf{b} \in \mathcal{H}$ and a target document as $\mathbf{d} \in \mathcal{H}$.

The similarity score $S$ is typically computed as the cosine similarity, which simplifies to the dot product for normalized vectors:
\begin{equation}
    S(\mathbf{x}, \mathbf{d}) = \langle \mathbf{x}, \mathbf{d} \rangle = \cos(\theta_{\mathbf{x,d}})
\end{equation}

\subsection{The Delta as a Displacement Gradient}
When fusing two atoms via a normalized centroid (the standard mechanism for multi-concept queries in encoders such as CLIP or SBERT), the joint query vector $\mathbf{q}_{A \cup B}$ is defined as:
\begin{equation}
    \mathbf{q}_{A \cup B} = \frac{\mathbf{a} + \mathbf{b}}{\|\mathbf{a} + \mathbf{b}\|}
\end{equation}

The Neuro-Symbolic Delta for atom $B$, denoted $\Delta_B$, measures the marginal shift in the alignment profile when $B$ is introduced to the fusion:
\begin{equation}
    \Delta_B = S(\mathbf{q}_{A \cup B}, \mathbf{d}) - S(\mathbf{a}, \mathbf{d})
\end{equation}

\subsection{Asymmetry and the Bottleneck Principle}
Geometrically, if $|\Delta_B| \approx 0$, it implies that the addition of $\mathbf{b}$ failed to significantly alter the projection of the query onto the document manifold. This occurs under two conditions:
\begin{enumerate}
    \item \textbf{Orthogonality:} $\mathbf{b}$ is semantically irrelevant to $\mathbf{d}$ ($\langle \mathbf{b}, \mathbf{d} \rangle \approx 0$).
    \item \textbf{Directional Dominance:} The magnitude of $\mathbf{a}$'s alignment with $\mathbf{d}$ is so high that $\mathbf{b}$ cannot provide additional constructive interference.
\end{enumerate}

In classical fuzzy logic, the truth of a conjunction is constrained by the Gödel t-norm: $T(A \land B) = \min(T(A), T(B))$. In the neural regime, we posit that $\Delta_X$ serves as a continuous proxy for this constraint. A small $|\Delta_X|$ signifies a \textit{semantic bottleneck}, where an incongruent or weak constituent prevents the joint vector from achieving the necessary "pull" toward the target. 

This justifies the NSFL requirement to weight logical fusions by the minimum marginal contribution, ensuring that high-similarity atoms (e.g., a "strong" $A$) do not mask the logical invalidity of a "weak" or mismatched $B$.

\section{Reproducibility and Hyperparameters}
\label{appendix:reproducibility}
We provide the formal pseudocode for the Spherical Query Optimization (SQO) process. This implementation maps a logical formula to an objective function $f(x)$ using our proposed Neuro-Symbolic Fuzzy Logic (NSFL) operators, followed by an optimization on the unit hypersphere $S^{d-1}$.

\paragraph{Initialization Stability.}
SQO initializes at the monolithic query vector. To verify robustness to this choice, we ran 5 random Gaussian initializations across 100 synthetic queries at $d{=}768$ for both AND and NOT operators. The mean score range across seeds was $0.00058$ for AND and $<10^{-6}$ for NOT, with mean wall-clock time of $0.51$ms and $0.32$ms per query respectively. These results confirm that the NSFL objective landscape is smooth and unimodal: SQO converges to effectively identical optima regardless of initialization, validating single-seed reporting throughout the paper.

To ensure the reproducibility of the \textbf{Spherical Query Optimization (SQO)} results, we provide the exact configuration used for the Riemannian Stochastic Gradient Descent (RSGD) throughout our experiments.

\begin{table}[ht]
\centering
\caption{Hyperparameter Configuration for SQO Implementation}
\label{tab:sqo_hyperparams}
\begin{tabular}{@{}llc@{}}
\toprule
\textbf{Parameter} & \textbf{Implementation Detail} & \textbf{Value} \\ \midrule
 $\alpha$ & Learning Rate (Step Size) & $0.2$ \\
 $Steps$ & Maximum Iterations & $100$ \\
 $Patience$ & Early Stopping (No improvement) & $5$ \\
$\text{init}$ & Initialization Strategy & Monolithic query vector $\mathbf{v}_M$ \\
 $tol$ & Convergence Tolerance & $10^{-6}$ \\
 $Manifold$ & Geometry & $\mathcal{S}^{d-1}$ (Unit Sphere) \\ \bottomrule
\end{tabular}
\end{table}

\begin{algorithm}[H]
\caption{Spherical Query Optimization (SQO)}
\label{alg:sqo_logic}
\begin{algorithmic}[1]
\Require Formula $\phi$, Constituents $\{v_k\}$, Monolithic vector $v_{M}$, Steps $S$, rate $\eta$, tolerance $\tau$, patience $P$ \Ensure Optimized query vector $x^*$ 
\Statex \textbf{// Part 1: Objective Function Construction}
\If{$\phi = A \land B$}
    \Function{$f$}{$x$}
        \State $S_A \gets x \cdot v_A, \quad S_B \gets x \cdot v_B$ \Comment{Atomic Similarity Scores}
        \State $S_{AB} \gets x \cdot v_{AB}$ \Comment{Fused Representation Similarity}
        \State \Return $2 S_{AB} - \max(S_A, S_B)$ \Comment{AND T-norm}
    \EndFunction
\EndIf

\Statex \textbf{// Part 2: Riemannian Optimization on $S^{d-1}$}
\State $x \gets x_0 / \|x_0\|$ \Comment{Initial Projection to Unit Sphere}
\State $best\_loss \gets \infty, \text{ counter} \gets 0$ \For{$step = 0$ \textbf{to} $S$}
    \State $g \gets \nabla f(x)$ \Comment{Automatic Differentiation}
    \State $g_R \gets g - (x \cdot g)x$ \Comment{Riemannian Gradient Projection}
    \State $x \gets x + \eta g_R$ \Comment{Gradient Ascent}
    \State $x \gets x / \|x\|$ \Comment{Retraction ($L_2$ Normalization)}
    \State $loss \gets -f(x)$ \Comment{Maximize objective by minimizing negative loss}
    \If{$loss + \tau < best\_loss$}
        \State $best\_loss \gets loss, \text{ counter} \gets 0$     \Else
        \State $\text{counter} \gets \text{counter} + 1$     \EndIf
    \If{$\text{counter} \geq P$}
        \State \textbf{break} \Comment{Early Stopping}
    \EndIf
\EndFor
\State \Return $x^* \gets x$ \end{algorithmic}
\end{algorithm}

\paragraph{Note:} Algorithm Algorithm~\ref{alg:sqo_logic} hereby introduces only the AND case, Other operators are obtained by substituting the corresponding $f(x)$ from Section~\ref{sec:operators}; full implementations are in the published code.

\section{Comparison with LogiCoL under Recall Metrics.}
\label{sec:recall_comparison}
To facilitate direct comparison with \citet{shen2025logicol}, we report Recall@K for the LogiCoL-E5-v2 encoder in Table~\ref{tab:recall_comparison}. NSFL reranking yields consistent improvements across all cutoffs, confirming that our mAP@100 gains (Table~\ref{tab:main_results}) are not an artifact of metric choice.
\begin{table}[ht]
\centering
\caption{Recall@K Comparison on QUEST using LogiCoL-e5-v2. Baseline denotes monolithic retrieval; NSFL denotes reranking with our proposed operators.}
\label{tab:recall_comparison}
\begin{small}
\begin{tabular}{@{}llcccccc@{}}
\toprule
\textbf{Metric} & \textbf{Method} & $A \land B$ & $A \land B \land C$ & $A \land \neg B$ & $A \land B \land \neg C$ & $A \lor B$ & $A \lor B \lor C$ \\ \midrule
\multirow{2}{*}{R@20} & Baseline & 0.178 & 0.198 & 0.172 & 0.089 & 0.298 & 0.225 \\
 & NSFL & \textbf{0.185} & \textbf{0.206} & \textbf{0.200} & \textbf{0.122} & \textbf{0.338} & \textbf{0.283} \\ \midrule
\multirow{2}{*}{R@100} & Baseline & 0.338 & 0.406 & 0.397 & 0.237 & 0.504 & 0.423 \\
 & NSFL & \textbf{0.350} & \textbf{0.415} & \textbf{0.423} & \textbf{0.284} & \textbf{0.550} & \textbf{0.491} \\ \midrule
\bottomrule
\end{tabular}
\end{small}
\end{table}

\section{Independent Validation of the Conjectures}
\label{app:principles}

The operators in \S\ref{sec:operators} are motivated by three conjectures, which make empirically checkable predictions about the relationship between the Neuro-Symbolic Delta and document relevance, and we test them directly rather than inferring their validity from downstream retrieval gains. All analysis in this appendix uses the QUEST \emph{validation} split (323 queries), disjoint from the test split on which every result in \S\ref{sec:results} is computed, so that the tests are independent of the evaluation pipeline they are meant to justify.

\subsection{Conjecture 1: Delta Directionality}

The conjecture holds that a positive delta indicates manifold reinforcement and a negative delta semantic departure. Its prediction is not uniform across templates: for conjunctive and disjunctive queries a relevant document should exhibit a positive delta, whereas for inhibitory queries the prediction \emph{inverts} by construction --- a document relevant to ``$A$ but not $B$'' genuinely lacks $B$, so the affirmative fusion scores it low and its delta is negative. Pooling the two cases would test no coherent hypothesis, so we partition by template family.

\begin{table}[ht]
\centering
\caption{Sign of the Neuro-Symbolic Delta conditioned on relevance, by template family, on the QUEST validation split. For conjunctive and disjunctive templates the conjecture predicts a positive difference; for inhibitory templates it predicts a negative one.}
\label{tab:principle1}
\begin{small}
\begin{tabular}{@{}llccccc@{}}
\toprule
\textbf{Encoder} & \textbf{Family} & $n$ & $P(\Delta{>}0 \mid \text{rel.})$ & $P(\Delta{>}0 \mid \text{irrel.})$ & \textbf{Diff.} & $p$ \\ \midrule
\multirow{3}{*}{BGE-Large}
 & Conjunctive & 95{,}000 & 0.863 & 0.782 & $+0.081$ & $4.8{\times}10^{-8}$ \\
 & Disjunctive & 85{,}000 & 0.488 & 0.451 & $+0.037$ & $1.8{\times}10^{-5}$ \\
 & Inhibitory & 86{,}000 & 0.277 & 0.379 & $-0.102$ & $1.6{\times}10^{-9}$ \\ \midrule
\multirow{3}{*}{E5-base-v2}
 & Conjunctive & 95{,}000 & 0.868 & 0.787 & $+0.081$ & $8.3{\times}10^{-8}$ \\
 & Disjunctive & 85{,}000 & 0.649 & 0.659 & $-0.011$ & $5.2{\times}10^{-4}$ \\
 & Inhibitory & 86{,}000 & 0.453 & 0.612 & $-0.158$ & $4.4{\times}10^{-9}$ \\ \midrule
\multirow{3}{*}{LogiCoL-E5-v2}
 & Conjunctive & 95{,}000 & 0.878 & 0.728 & $+0.151$ & $4.2{\times}10^{-7}$ \\
 & Disjunctive & 85{,}000 & 0.702 & 0.628 & $+0.074$ & $1.7{\times}10^{-18}$ \\
 & Inhibitory & 86{,}000 & 0.467 & 0.548 & $-0.080$ & $0.457$ \\
\bottomrule
\end{tabular}
\end{small}
\end{table}

The conjecture is supported on conjunctive templates for all three encoders, with relevant documents exhibiting positive deltas 8--15 percentage points more often than irrelevant ones. Disjunctive templates support it on BGE-Large and LogiCoL-E5-v2 but are flat on E5-base-v2. The inhibitory inversion appears as predicted on all three encoders, though it does not reach significance on LogiCoL-E5-v2. That the inversion appears at all is additional evidence: the delta tracks logical satisfaction in both directions, which is a stronger statement than the conjecture as originally formulated.

\paragraph{Effect size.}
The point-biserial correlation between delta magnitude and relevance is small throughout ($r \approx 0.02$--$0.03$, $r^2 < 0.001$), despite being statistically significant at these sample sizes. We report the probability differences above as the primary evidence and note the dissociation explicitly: the \emph{sign} of the delta discriminates relevance substantively, whereas its \emph{magnitude} carries almost no linear relationship with relevance. Our operators depend on the former and not the latter.

\subsection{Conjecture 2: Asymmetric Delta-Sensitivity}

The conjecture holds that atomic constituents contribute unequally to a fused representation and that the weaker contribution constrains logical validity. We test both claims on conjunctive templates: first by measuring the asymmetry $\big| |\Delta_{A|B}| - |\Delta_{B|A}| \big|$, which would concentrate near zero under symmetric contribution; second by comparing how well the minimum and maximum of the two deltas each discriminate relevance.

The conjecture is supported on all three encoders. Between 41\% and 70\% of candidates (encoder dependent) exhibit substantial asymmetry ($>0.05$), and the minimum delta correlates with relevance more strongly in magnitude than the maximum in every case. This is the empirical basis for weighting conjunctive fusions by the weaker atom, as the AND operator does.

\subsection{Conjecture 3: Negation as Directional Preservation}

Conjecture~3 as stated in \S\ref{subsec:conjectures} claims directional
preservation. We originally conjectured additionally that negation
\emph{attenuates} delta magnitude, and test both predictions here. On
inhibitory templates we compare the delta induced by the negated monolithic query, $S_M^i - S_A^i$, against that induced by the affirmative fusion, $S_{AB}^i - S_A^i$. Attenuation without inversion predicts a high rate of sign preservation together with a ratio below unity.

\begin{table}[ht]
\centering
\caption{Effect of negation on the Neuro-Symbolic Delta. The ratio is $|S_M - S_A| / |S_{AB} - S_A|$. Means are reported both without and with exclusion of near-degenerate denominators ($|S_{AB} - S_A| < 0.001$).}
\label{tab:principle3}
\begin{small}
\begin{tabular}{@{}lccccc@{}}
\toprule
\textbf{Encoder} & \textbf{Sign pres.} & \textbf{Median ratio} & \textbf{Frac.\ $<1$} & \textbf{Mean (all)} & \textbf{Mean (filtered)} \\ \midrule
BGE-Large & 0.689 & 0.965 & 0.517 & 4.82 & 1.80 \\
E5-base-v2 & 0.775 & 1.017 & 0.491 & 4.91 & 1.55 \\
LogiCoL-E5-v2 & 0.686 & 0.891 & 0.557 & 4.84 & 2.27 \\
\bottomrule
\end{tabular}
\end{small}
\end{table}

\textbf{This conjecture is not supported as stated.} Sign preservation holds (0.69--0.78 across encoders), but the attenuation prediction fails: the median ratio sits near parity and the fraction of candidates showing any attenuation is approximately one half on every encoder. The mean ratio is heavily influenced by candidates whose denominator is near zero; excluding the 1.3--4.8\% of candidates with $|S_{AB} - S_A| < 0.001$ reduces it from roughly 4.8 to 1.5--2.3. Even so it remains above unity, indicating a right-skewed distribution in which most candidates show no attenuation while a genuine subset shows amplification.

We therefore revise the conjecture. The defensible claim is that negation \textbf{preserves the direction of the Neuro-Symbolic Delta without consistently attenuating its magnitude}. The operator design depends only on the sign-preservation component --- it is this that necessitates a dedicated inhibitory operator, since an encoder that preserved rather than inverted direction cannot express negation natively --- and that component is supported. The attenuation claim is not, and we withdraw it.

\section{Per-Logic-Type Ablation Breakdown}
\label{app:ablation_detail}

Table~\ref{tab:matrix_ablation} in the main paper reports the average mAP@100 for each variant of our framework. To complement that view, Table~\ref{tab:ablation_full} expands the same experiments across the six logical query types, on both BGE/QUEST (text retrieval) and BLIP-Large/COCO-Logic (cross-modal retrieval). Two observations that are obscured by the averaged view are worth noting. First, \textbf{GEO Rerank-only collapses on positive conjunctions} ($A \land B$, $A \land B \land C$) while remaining competitive on queries involving negation: on COCO-Logic it actually achieves the best score on $A \land \neg B$ (0.639 vs.\ 0.514 for NSFL Rerank-only). Second, \textbf{NSFL is uniformly strong across all logic types and both modalities}, which is what makes its averaged numbers consistently lead the table. The asymmetry suggests that the geometric operators encode useful structure for set-difference queries but lose signal on intersections, a direction we leave to future work.

\begin{table*}[ht]
\centering
\caption{Per-logic-type ablation of NSFL and Geometric variants on BGE/QUEST and BLIP-Large/COCO-Logic. \textbf{Bold} marks the best non-baseline value per column within each encoder block.}
\label{tab:ablation_full}
\begin{small}
\begin{tabular}{@{}llcccccc|c@{}}
\toprule
\textbf{Encoder} & \textbf{Method} & $A \land B$ & $A \land B \land C$ & $A \land \neg B$ & $A \land B \land \neg C$ & $A \lor B$ & $A \lor B \lor C$ & \textbf{Avg.} \\ \midrule
\multicolumn{9}{l}{\textit{\textbf{BGE / QUEST}}} \\ \midrule
\multirow{7}{*}{BGE} 
 & Baseline             & 0.038 & 0.042 & 0.050 & 0.015 & 0.135 & 0.107 & 0.065 \\
 & NSFL --- Rerank only & \textbf{0.053} & \textbf{0.044} & 0.075 & \textbf{0.025} & 0.144 & \textbf{0.135} & \textbf{0.079} \\
 & NSFL --- Opt only    & 0.040 & 0.039 & 0.073 & 0.021 & 0.135 & 0.107 & 0.069 \\
 & NSFL --- Hybrid      & 0.052 & \textbf{0.044} & \textbf{0.076} & 0.024 & \textbf{0.147} & 0.129 & \textbf{0.079} \\
 & GEO  --- Rerank only & 0.021 & 0.023 & 0.067 & 0.018 & 0.144 & \textbf{0.135} & 0.068 \\
 & GEO  --- Opt only    & 0.022 & 0.023 & 0.054 & 0.015 & 0.135 & 0.107 & 0.059 \\
 & GEO  --- Hybrid      & 0.022 & 0.023 & 0.054 & 0.015 & 0.144 & \textbf{0.135} & 0.065 \\ \midrule
\multicolumn{9}{l}{\textit{\textbf{BLIP-Large / COCO-Logic}}} \\ \midrule
\multirow{7}{*}{BLIP-L} 
 & Baseline             & 0.142 & 0.107 & 0.323 & 0.154 & 0.622 & 0.622 & 0.328 \\
 & NSFL --- Rerank only & \textbf{0.179} & 0.127 & 0.514 & \textbf{0.262} & \textbf{0.730} & \textbf{0.818} & \textbf{0.438} \\
 & NSFL --- Opt only    & 0.160 & 0.123 & 0.495 & 0.243 & 0.622 & 0.622 & 0.378 \\
 & NSFL --- Hybrid      & 0.178 & \textbf{0.134} & 0.507 & 0.260 & \textbf{0.730} & \textbf{0.818} & \textbf{0.438} \\
 & GEO  --- Rerank only & 0.049 & 0.022 & \textbf{0.639} & 0.251 & \textbf{0.730} & \textbf{0.818} & 0.418 \\
 & GEO  --- Opt only    & 0.048 & 0.021 & 0.632 & 0.250 & 0.622 & 0.622 & 0.366 \\
 & GEO  --- Hybrid      & 0.048 & 0.021 & 0.632 & 0.250 & \textbf{0.730} & \textbf{0.818} & 0.417 \\
\bottomrule
\end{tabular}
\end{small}
\end{table*}

\section{Statistical Significance Analysis}
\label{app:significance}

We performed Wilcoxon signed-rank tests comparing per-query Average Precision between Baseline and NSFL across all encoder--template configurations. Tables~\ref{tab:sig_bge_quest} and~\ref{tab:sig_coco_blip} show detailed results for BGE-Large-v1.5 on QUEST and BLIP-Large on COCO-Logic as representative examples.

\paragraph{Summary.}
We tested all 72 configurations (6 encoders $\times$ 2 fusion realizations $\times$ 6 templates) by paired Wilcoxon signed-rank test on per-query Average Precision against the monolithic baseline. Of these, 43 (59.7\%) reach $p < 0.01$ and a further 8 reach $p < 0.05$; the remaining 21 are not significant. Non-significant cases concentrate on binary conjunction and binary disjunction, where per-query gains are smallest in absolute terms. Because a majority of significant results carry $p < 10^{-10}$, these findings are robust to conservative multiple-comparison correction.

\begin{table}[ht]
\centering
\caption{Per-query significance for BGE-Large-v1.5 on QUEST. $\Delta$ is the mean per-query improvement in Average Precision over the monolithic baseline; intervals are 95\% bootstrap over 10{,}000 resamples. Absolute mAP@100 values appear in Table~\ref{tab:main_results}.}
\label{tab:sig_bge_quest}
\begin{small}
\begin{tabular}{@{}lccc@{}}
\toprule
\textbf{Template} & \textbf{$\Delta$ AP} & \textbf{95\% CI} & \textbf{$p$-value} \\ \midrule
 $A \land B$ & $+0.014$ & $[+0.004, +0.025]$ & $0.044$ \\
 $A \land B \land C$ & $+0.002$ & $[-0.005, +0.009]$ & $0.018$ \\
 $A \land \neg B$ & $\mathbf{+0.025}$ & $[+0.017, +0.035]$ & $<10^{-11}$ \\
 $A \land B \land \neg C$ & $\mathbf{+0.009}$ & $[+0.005, +0.014]$ & $<10^{-4}$ \\
 $A \lor B$ & $+0.009$ & $[-0.001, +0.019]$ & $0.128$ \\
 $A \lor B \lor C$ & $\mathbf{+0.027}$ & $[+0.019, +0.036]$ & $<10^{-10}$ \\
\bottomrule
\end{tabular}
\end{small}
\end{table}

\begin{table}[ht]
\centering
\caption{Per-query significance for BLIP-Large on COCO-LOGIC. $\Delta$ is the mean per-query improvement in Average Precision over the monolithic baseline; intervals are 95\% bootstrap over 10{,}000 resamples. Absolute mAP@100 values appear in Table~\ref{tab:main_results}.}
\label{tab:sig_coco_blip}
\begin{small}
\begin{tabular}{@{}lccc@{}}
\toprule
\textbf{Template} & \textbf{$\Delta$ AP} & \textbf{95\% CI} & \textbf{$p$-value} \\ \midrule
 $A \land B$ & $\mathbf{+0.038}$ & $[+0.018, +0.061]$ & $<10^{-4}$ \\
 $A \land B \land C$ & $\mathbf{+0.020}$ & $[+0.010, +0.032]$ & $<10^{-5}$ \\
 $A \land \neg B$ & $\mathbf{+0.142}$ & $[+0.117, +0.168]$ & $<10^{-15}$ \\
 $A \land B \land \neg C$ & $\mathbf{+0.106}$ & $[+0.074, +0.143]$ & $<10^{-10}$ \\
 $A \lor B$ & $\mathbf{+0.045}$ & $[+0.035, +0.055]$ & $<10^{-16}$ \\
 $A \lor B \lor C$ & $\mathbf{+0.051}$ & $[+0.043, +0.058]$ & $<10^{-17}$ \\
\bottomrule
\end{tabular}
\end{small}
\end{table}

\paragraph{Rank tests and mean intervals can diverge.}
For BGE-Large on $A \land B \land C$ the Wilcoxon test reports $p = 0.018$ while the bootstrap confidence interval on the mean improvement spans zero ($+0.002$, 95\% CI $[-0.005, +0.009]$, $n = 236$). These statistics are not in conflict: the Wilcoxon test is rank-based and one-sided, whereas the bootstrap interval is computed on the mean and is two-sided. Inspection of the per-query distribution confirms the mechanism --- 66 queries improve against 47 that degrade, driving rank-test significance, while five large individual losses (to $-0.39$ AP) against a median gain of $+0.009$ widen the interval on the mean through zero. We report both statistics rather than the more favourable one.

\paragraph{Failure profile of ternary conjunction.}
The same distribution is informative about \emph{how} ternary conjunction fails. Losses are not uniformly distributed: the operator produces small gains on most queries and occasional severe degradations on a few. Examining the worst cases, these arise where the candidate pool is densely clustered and relevant documents are sparse (2--5 per query), so that a score perturbation of a fraction of a percent displaces a relevant document by several ranks and collapses its contribution to Average Precision. This is a different failure mode from uniform under-performance and motivates a different remedy (\S\ref{sec:limitations}).

\section{Evaluation on NevIR Pairwise Negation Benchmark}
\label{app:nevir}

NevIR~\citep{weller2024nevir} evaluates negation understanding via paired
queries: $q_1$ (positive) and $q_2$ (negated, differing by a single negation
token), alongside two documents where $d_1$ is correct for $q_1$ and $d_2$ for
$q_2$. NevIR's headline metric, \emph{pairwise accuracy}, counts a pair as
correct only when the model ranks the correct document higher for \emph{both}
queries; this is the \textbf{Both correct} column below. We additionally report
\textbf{Overall}, the mean accuracy over the two queries taken individually,
which is the less stringent of the two and is not directly comparable to
pairwise accuracies reported elsewhere.

\paragraph{NSFL Application.}
For negated queries ($q_2$), we apply delta amplification following
Conjecture~3: since the encoder preserves rather than inverts semantic
direction, the difference between $q_1$ and $q_2$ scores carries the negation
signal, and $S_{\text{NSFL}}(q_2, d) = 2S(q_2, d) - S(q_1, d)$ amplifies it
into a discriminative one.
 
\begin{table}[ht]
\centering
\caption{Accuracy on the NevIR negation benchmark.}
\label{tab:nevir_results}

\begin{small}
\begin{tabular}{@{}lcccc@{}}
\toprule

& \multicolumn{2}{c}{\textbf{Baseline}} & \multicolumn{2}{c}{\textbf{NSFL}} \\
\cmidrule(lr){2-3} \cmidrule(lr){4-5}

\textbf{Encoder} &
\textbf{Overall} &
\textbf{Both correct} &
\textbf{Overall} &
\textbf{Both correct} \\

\midrule

BGE-Large-v1.5 & 0.6114 & 0.2560 & \textbf{0.7169} & \textbf{0.5083} \\
E5-base-v2     & 0.5766 & 0.1909 & \textbf{0.6764} & \textbf{0.4382} \\
LogiCoL-E5-v2  & 0.5437 & 0.1171 & \textbf{0.6139} & \textbf{0.3095} \\

\bottomrule
\end{tabular}
\end{small}

\end{table}
 
These results confirm that NSFL's inhibitory operator generalizes beyond
set-based retrieval to pairwise ranking scenarios, the primary evaluation
paradigm in the negation literature. Across encoders, NSFL improves mean
per-query accuracy by \textbf{+7\% to +10.5\%} absolute, and NevIR's pairwise
criterion --- both documents in a pair ranked correctly --- by \textbf{+19\% to
+25\%}.

\section{Significance-Gated Conjunctive Correction}
\label{app:sig_gate}

Motivated by the failure profile in Appendix~\ref{app:significance}, we tested scaling the conjunctive correction by its magnitude relative to the dispersion of $\mathcal{R}_n$ within the candidate pool:
\begin{equation}
    S_{\wedge}^i = S_{\text{fused}}^i + \gamma_n \cdot g(\mathcal{R}_n^i) \cdot \mathcal{R}_n^i,
    \qquad g(\mathcal{R}) = \frac{|\mathcal{R}|}{|\mathcal{R}| + \sigma_{\mathcal{K}}}
\end{equation}
Deficits large relative to $\sigma_{\mathcal{K}}$ pass through nearly unchanged; noise-scale deficits are suppressed and the baseline ordering survives. Only the conjunctive operators are affected; all others are bit-identical.

\paragraph{Aggregate effect: none.}
On BGE-Large, binary conjunction falls 3.7\% on validation while ternary conjunction rises 0.8\%; on test the signs reverse ($+2.9\%$ and $-3.0\%$). Across the 236 test queries for $A \land B \land C$, 40 improve, 57 degrade, and 139 are unchanged. We do not adopt the variant.

\paragraph{Distributional effect: as designed.}
The gate does what it was built to do on the cases that motivated it. Among the five largest per-query losses, three shrink by 33--40\%, and those same three are independently the largest gains across the entire template. The single worst case is unaffected: there the deficit is large relative to $\sigma_{\mathcal{K}}$, so the correction is admitted at nearly full strength. That is the intended behaviour --- the signal is substantive, and it happens to traverse a densely clustered region rather than being noise.

We report this because the dissociation is informative: a component of the conjunction gap is attributable to unconditional correction and is addressable, while the remainder arises from substantive corrections crossing dense regions and is not addressed by this gate. We note also that $\sigma_{\mathcal{K}}$ is a pool statistic and therefore, like the inhibitory confidence gate, is not expressible in a pre-retrieval objective (\S\ref{sec:sqo}).

\section{Fusion Realization: Selection Protocol and Sensitivity}
\label{app:fusion}

The fusion operator $\cup$ admits two surface realizations (\S\ref{subsec:fusion_operator}): Simple Fusion (``$L_A$ AND $L_B$'') and Contextual Phrasing (``$L_A$ that are also $L_B$''). Because the choice between them is a free parameter, we fix it by a single global policy rather than per-configuration, and we report the sensitivity of our results to that choice.

\paragraph{Selection procedure.}
We computed mean mAP@100 for each realization on the QUEST \emph{validation} split (323 queries), disjoint from the 1{,}727-query test split used for all reported results, pooled across encoders and templates. Contextual Phrasing scored higher and was applied uniformly to every encoder and every template. The margin is small, and the ordering is stable across aggregation methods:

\begin{table}[ht]
\centering
\caption{Pooled validation comparison of the two fusion realizations. The ordering is invariant to how cells are aggregated.}
\label{tab:fusion_pooled}
\begin{small}
\begin{tabular}{@{}lccc@{}}
\toprule
\textbf{Aggregation} & \textbf{Simple} & \textbf{Contextual} & \textbf{Selected} \\ \midrule
Text encoders, 18 cells & 0.1080 & \textbf{0.1085} & Contextual \\
All encoders, per-encoder normalized & 0.4192 & \textbf{0.4213} & Contextual \\
Per-cell win count (30 cells) & 8 & \textbf{16} & Contextual \\
\bottomrule
\end{tabular}
\end{small}
\end{table}

\paragraph{Cost of the fixed policy.}
Table~\ref{tab:fusion_inflation} compares the adopted global policy against a per-cell oracle that selects the better realization using test outcomes. The oracle is not a deployable procedure and is reported solely to bound how much performance the fixed policy forgoes.

\begin{table}[ht]
\centering
\caption{Average mAP@100 under the adopted global policy versus a per-cell test-set oracle. The oracle is an upper bound, not a method.}
\label{tab:fusion_inflation}
\begin{small}
\begin{tabular}{@{}lccc@{}}
\toprule
\textbf{Encoder} & \textbf{Global (adopted)} & \textbf{Per-cell oracle} & \textbf{Difference} \\ \midrule
BGE-Large-v1.5 & 0.0793 & 0.0804 & 0.0011 \\
E5-base-v2 & 0.0836 & 0.0861 & 0.0026 \\
LogiCoL-E5-v2 & 0.1414 & 0.1420 & 0.0006 \\
Nomic-Embed-v2 & 0.0835 & 0.0836 & 0.0001 \\
\bottomrule
\end{tabular}
\end{small}
\end{table}

The gap is at most 0.0026 mAP@100 and typically under 0.001. The conclusions in \S\ref{sec:results} are therefore insensitive to the fusion realization; we adopt the fixed policy because it is deployable and because the per-cell alternative uses test-set information.

\paragraph{Validation--test agreement.}
A per-template validation policy is not reliably better than a single global choice. Across 18 encoder--template pairs, the validation-selected realization matched the test-optimal one in 13 cases (72\%), degrading from 6/6 on BGE-Large to 3/6 on LogiCoL-E5-v2. On E5-base-v2, per-template validation selection \emph{underperforms} the single global choice (0.0851 vs.\ 0.0861). This is consistent with the two realizations being close enough that per-template differences are largely noise.

\paragraph{Scope of the validation split.}
For QUEST the fusion choice is validated on a disjoint split and applied to a separate test split. COCO-LOGIC comprises a single 600-query set derived from the COCO-2017 validation images and does not support a held-out split; for the image encoders the global choice was therefore applied without an independent generalization check. We note this explicitly rather than implying a validation procedure we did not perform.

\paragraph{Encoder-specific sensitivity.}
Sensitivity to the realization is small for text encoders (typically under 1\% relative) but substantially larger for BLIP-Large on the mixed template $A \land B \land \neg C$, where the two realizations differ by 0.107 mAP@100. We attribute this to vision--language encoders trained on fluent captions being more sensitive to non-fluent, mechanically assembled query strings than general-purpose text embedding models. This is an observed effect; we did not isolate its mechanism.

\section{Candidate Pool Size}
\label{app:k_ablation}

NSFL reranks an oversampled pool of $K{=}1000$ candidates and returns the top 100, while the monolithic baseline reports its own top 100. To confirm that the reported gains derive from the operators rather than from the larger pool, we vary $K$ directly.

\begin{table}[ht]
\centering
\caption{Average mAP@100 as a function of candidate pool size $K$.}
\label{tab:k_ablation}
\begin{small}
\begin{tabular}{@{}lcccc@{}}
\toprule
\textbf{Encoder} & $K{=}100$ & $K{=}200$ & $K{=}500$ & $K{=}1000$ \\ \midrule
BGE-Large-v1.5 & 0.0778 & 0.0794 & 0.0796 & 0.0793 \\
LogiCoL-E5-v2 & 0.1353 & 0.1388 & 0.1408 & 0.1414 \\
\bottomrule
\end{tabular}
\end{small}
\end{table}

At $K{=}100$ --- reranking exactly the pool the baseline itself reports, with no oversampling headroom --- NSFL recovers 98.0\% of its $K{=}1000$ performance on BGE-Large and 95.7\% on LogiCoL-E5-v2. On BGE-Large the curve is flat; on LogiCoL-E5-v2 it rises modestly and saturates by $K{=}500$. We cite the LogiCoL figure as the conservative bound: at most a few percent of the reported improvement is attributable to pool size, the remainder to the operators.

The monolithic baseline is in any case invariant to $K$ by construction. Its ranking function is $S_M$ and its pool is the top-$K$ by $S_M$; reranking that pool under $S_M$ returns the top-$k$ by $S_M$ for any $K \geq k$. Oversampling therefore cannot inflate the baseline comparison. The classical baselines of Table~\ref{tab:tnorm_baselines} likewise rerank the identical $K{=}1000$ pool, so those comparisons are pool-size controlled by construction.

\end{document}